\documentclass[journal]{IEEEtran}

\usepackage{amssymb,amsmath,graphicx,times,cite,multirow,color,subfigure,url}
\usepackage[lined,ruled,boxed]{algorithm2e}

\begin{document}
\title{Different Set Domain Adaptation for Brain- Computer Interfaces: A Label Alignment Approach}

\author{He~He and Dongrui~Wu
\thanks{He~He and Dongrui~Wu are with the Ministry of Education Key Laboratory of Image Processing and Intelligent Control, School of Artificial Intelligence and Automation, Huazhong University of Science and Technology, Wuhan, China. Email: hehe91@hust.edu.cn, drwu@hust.edu.cn.}
\thanks{Dongrui Wu is the corresponding author.}}

\maketitle

\begin{abstract}
A brain-computer interface (BCI) system usually needs a long calibration session for each new subject/task to adjust its parameters, which impedes its transition from the laboratory to real-world applications. Domain adaptation, which leverages labeled data from auxiliary subjects/tasks (source domains), has demonstrated its effectiveness in reducing such calibration effort. Currently, most domain adaptation approaches require the source domains to have the same feature space and label space as the target domain, which limits their applications, as the auxiliary data may have different feature spaces and/or different label spaces. This paper considers different set domain adaptation for BCIs, i.e., the source and target domains have different label spaces. We introduce a practical setting of different label sets for BCIs, and propose a novel label alignment (LA) approach to align the source label space with the target label space. It has three desirable properties: 1) LA only needs as few as one labeled sample from each class of the target subject; 2) LA can be used as a preprocessing step before different feature extraction and classification algorithms; and, 3) LA can be integrated with other domain adaptation approaches to achieve even better performance. Experiments on two motor imagery datasets demonstrated the effectiveness of LA.
\end{abstract}

\begin{IEEEkeywords}
Brain-computer interface, EEG, label alignment, Riemannian geometry, domain adaptation, transfer learning
\end{IEEEkeywords}

\IEEEpeerreviewmaketitle

\section{Introduction} \label{sect:introduction}

A brain-computer interface (BCI) system \cite{Wolpaw2002,Lance2012} acquires the brain signal, decodes it, and then translates it into control commands for external devices, so that a user can interact with his/her surroundings using thoughts directly, bypassing the normal pathway of peripheral nerves and muscles. Electroencephalogram (EEG) may be the most popular BCI input signal due to its convenience, safety, and low cost. The pipeline for decoding EEG signals usually involves:
\begin{enumerate}
\item \emph{Signal processing}, which includes band-pass filtering and spatial filtering. Bandpass filtering reduces interferences and noise such as muscle artifacts, eye blinks, and DC drift. Spatial filtering combines different EEG channels to increase the signal-to-noise ratio. Common spatial patterns (CSP) \cite{Koles1990, Muller1999, Ramoser2000, he2018spatial} may be the most frequently used spatial filtering approach.
\item \emph{Feature extraction}. Different features, e.g., time domain, frequency domain, time-frequency domain, Riemannian space, could be used.
\item \emph{Classification}. Popular classifiers include linear discriminant analysis (LDA) and support vector machine (SVM).
\end{enumerate}

Recently, Barachant \emph{et al.} \cite{Barachant2012} proposed a novel preprocessing and classification pipeline in the Riemannian space, which integrated spatial filtering and feature extraction into one single step. This Riemannian pipeline uses the covariance matrices of the EEG trials, which are symmetric positive definite and lie on a Riemannian manifold \cite{Yger2017}. The covariance matrices encode spatial information of the brain activities, which are useful in many BCI tasks. A popular classifier in the Riemannian space, minimum distance to mean \cite{Barachant2012}, treats the covariance matrices as points on the Riemannian manifold, and uses their Riemannian distances to the class mean for classification. Another more sophisticated approach maps the covariance matrices from the Riemannian space to a Euclidean tangent space (TS) around the Riemannian mean, where the Riemannian space covariance matrices are transformed into Euclidean space vectors, and then used in Euclidean space classifiers as features.

Motor imagery \cite{He2015} is one of the most frequently used paradigms of BCIs. It is based on the voluntary modulation of the sensorimotor rhythm, which does not need any external stimuli. The imagined movements of different body parts (e.g., hands, feet, and tongue) cause modulations of brain rhythms in the involved cortical areas. So, they can be distinguished by decoding such brain rhythm modulations, and used to control external devices such as powered exoskeletons, wheelchairs, and robots.

Motor imagery-based BCIs were originally designed to help those with neuromuscular impairments \cite{Pfurtscheller2008}. Recent research has extended its application scope to able-bodied users \cite{Nicolas-Alonso2012,Erp2012}. However, EEG signals are very weak, and easily contaminated by interferences and noise. Moreover, individual differences make it difficult, if not impossible, to build a generic machine learning model optimal for all subjects. Usually a calibration session is needed to collect some subject-specific data for a new subject, which is time-consuming and user-unfriendly.

Researchers have proposed many different approaches \cite{Jayaram2016, drwuTNSRE2016, drwuTFS2017, drwuTHMS2017, drwuSMC2017, drwuTLCSP2017, Kang2009, Lotte2010} to reduce this calibration effort. One of them is transfer learning \cite{Pan2010}, or domain adaptation (DA). Its main idea is to leverage the data from auxiliary subjects (called source subjects or source domains) to improve the learning performance for a new subject (called target subject or target domain). A popular idea in DA is to project the source domain and target domain data into low dimensional subspaces where the geometrical shift or/and distribution shift are reduced, such as joint distribution adaptation (JDA) \cite{long2013transfer}, joint geometrical and statistical alignment (JGSA) \cite{zhang2017joint}, and manifold embedded distribution alignment (MEDA) \cite{wang2018visual}. Computational intelligence techniques have also been used in transfer learning, as reviewed by Lu \emph{et al.} \cite{Lu2015Transfer}. In BCIs, Zanini \emph{et al.} \cite{Zanini2018} proposed a Riemannian geometry framework to align EEG covariance matrices from different subjects in the Riemannian space. Recently, we \cite{he2019transfer} proposed a Euclidean alignment (EA) approach, which can be used as a preprocessing step before many Euclidean space feature extraction and pattern recognition algorithms.

However, most existing DA approaches assume that the source domains have the same feature space and label space as the target domain, which may not hold in many real-world applications. There have been some heterogenous feature space DA approaches \cite{drwuTNSRE2016,day2017survey, liu2018unsupervised}, which address the problem that the source domains have different feature spaces from the target domain. For example, in BCIs, Wu \emph{et al.} \cite{drwuTNSRE2016} performed transfer learning for heterogenous feature spaces: the source and target EEG trials are collected from different EEG headsets, with different numbers of channels and channel locations. Its main idea is to select the source domain channels closest to the target domain channels.

\begin{figure}[htpb]\centering
\includegraphics[width=.7\linewidth,clip]{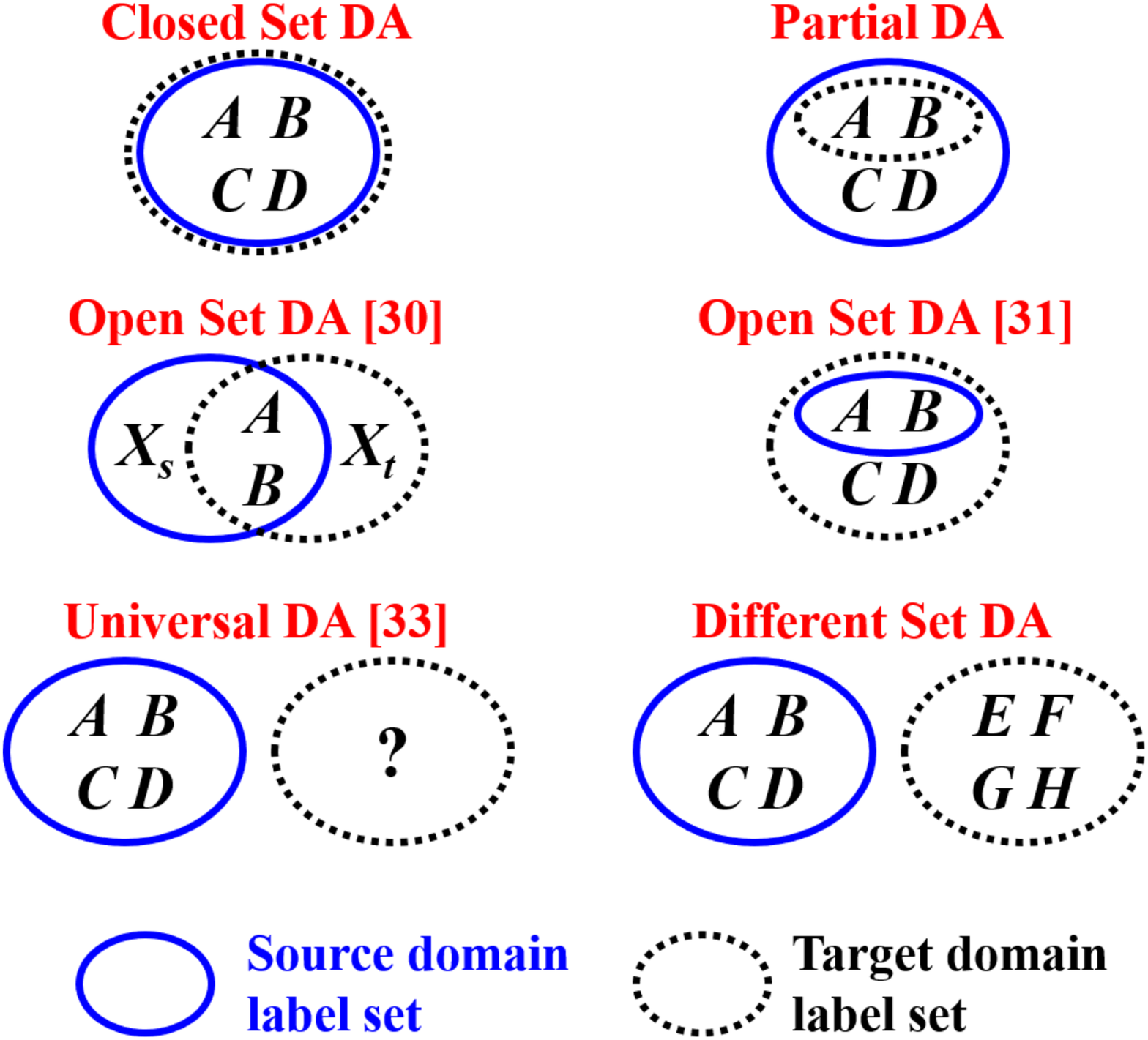}
\caption{Different DA scenarios. $A, B, ..., H$ represent different classes. $X_s$ and $X_t$ are unknown classes in the source domain and the target domain, respectively. In closed set DA, the source domain and target domain have the same classes. In partial DA, the target domain only contains a subset of the source domain classes. In open set DA considered in \cite{panareda2017open}, the source and target domains contain some common classes, but each also contains an ``unknown" class. In open set DA considered in \cite{saito2018open}, the source domain only contains a subset of the target domain classes. In universal DA \cite{you2019universal}, the target domain may contain some common classes as the target domain, but may also contain some unknown classes. In different set DA considered in this paper, the target domain contains partially or completely different classes from the source domain.} \label{fig:DA}
\end{figure}

There have also been a few heterogeneous label space DA approaches \cite{panareda2017open, saito2018open, fang2019open, you2019universal}, as shown in Fig.~\ref{fig:DA}. Busto \emph{et al.} \cite{panareda2017open} first proposed the concept of open set DA, assuming the source and target domains have some known classes in common, and also some classes that are different and unknown. Saito \emph{et al.} \cite{saito2018open} considered the case that the target domain contains all classes in the source domain, plus an ``unknown" class (different from \cite{panareda2017open}, herein the source domain does not contain an ``unknown" class). You \emph{et al.} \cite{you2019universal} proposed universal DA, which classifies a target domain sample if it belongs to any known class in the source domain, or marks it as ``unknown" otherwise. In summary, both open set DA and universal DA train a model to either classify a target domain sample into a known class which has appeared in the source domain, or mark it as ``unknown". An application scenario of open set DA and universal DA is face recognition, where some test samples may not appear in the training database and have to be marked as ``unknown".

This paper considers different set DA in BCIs, i.e., the source domains have different label spaces from the target domain, as shown in Fig.~\ref{fig:DA}. For Motor imagery-based BCIs, this means the source subjects and the target subject perform different motor imagery tasks. To our knowledge, no one has studied this problem before.

To address this issue, we propose a label alignment (LA) approach to align EEG covariance matrices of the source subjects to those of the target subject. It first matches each source domain label with a target domain label, then moves the per-class covariance matrices of each source subject to re-center them at the corresponding class means of the target subject. After LA, the distribution discrepancies between the source and the target subjects are reduced, so that a model trained on source subjects can classify each target trial into the category it actually belongs to, even though the source and target subjects have completely different label spaces.

The main contributions of this paper are:
\begin{enumerate}
  \item We introduce a practical setting of different set DA in BCIs: The source and target domains have known and different label sets; we need to classify each target trial into the category it actually belongs to, with the help of the source domain data. This setting is different from existing open set DA and universal DA. To our knowledge, it has not been studied before.
  \item We propose an effective LA approach for different set DA in BCIs, which has three desirable properties: 1) It only needs as few as one labeled EEG trial from each class of the target subject; 2) It can be used as a preprocessing step in different feature extraction and classification algorithms; and, 3) It can be integrated with other DA approaches to achieve even better performance.
\end{enumerate}

The remainder of this paper is organized as follows: Section~\ref{sect:RelatedWorks} introduces related background knowledge on the Riemannian space and the EA. Section~\ref{sect:LA} proposes the LA. Section~\ref{sect:datasets} introduces the datasets used in our experiments. Sections~\ref{sect:experiments} compares the performance of LA with several other DA approaches. Finally, Section~\ref{sect:conclusions} draws conclusions and points out some future research directions.

\section{Related Work} \label{sect:RelatedWorks}

This section introduces some basic concepts of the Riemannian space and its TS, and the EA, a state-of-the-art data alignment approach for BCIs, which also motivated our proposed LA.

\subsection{Riemannian Distance} \label{sect:RD}

Each symmetric positive definite matrix can be viewed as a point on a Riemannian manifold. The Riemannian distance between two symmetric positive definite matrices $P_1$ and $P_2$ is the length of the \textit{geodesic}, defined as the minimum length curve connecting $P_1$ and $P_2$ on the Riemannian manifold:
\begin{align}
 \delta(P_1,P_2)=\parallel \log(P_1^{-1}P_2)\parallel_F=\left[\sum_{r=1}^R\log^2\lambda_r\right]^{\frac{1}{2}}, \label{eq:geodesic}
\end{align}
where the subscript $F$ denotes the Frobenius norm, and $\lambda_r$ ($r=1,2,\ldots, R$) are the real eigenvalues of $P_1^{-1}P_2$.

$\delta(P_1,P_2)$ remains unchanged under linear invertible transformations:
\begin{align}
 \delta(W^TP_1W,W^TP_2W)= \delta(P_1,P_2), \label{eq:congruence invariance}
\end{align}
where $W$ is an invertible matrix. This property, called \emph{congruence invariance}, is useful in both EA and LA.

\subsection{Tangent Space (TS) Mapping}\label{sect:TS}

Most machine learning approaches are applicable only in the Euclidean space, and cannot be used in the Riemannian space. TS mapping maps the covariance matrices from the Riemannian space to a Euclidean TS, so that they can be used by a Euclidean space classifier.

For each point $P$ on the Riemannian manifold, the TS can be defined by a set of tangent vectors at $P$. Each tangent vector $S_i$ is defined as the derivative at $t=0$ of the geodesic between $P$ and the exponential mapping $P_i=\mbox{Exp}_P(S_i)$:
\begin{align}
\mbox{Exp}_P(S_i)=P_i=P^{\frac{1}{2}}\exp(P^{-\frac{1}{2}}S_iP^{-\frac{1}{2}})P^{\frac{1}{2}}.
\end{align}

The inverse mapping is given by the logarithmic mapping:
\begin{align}
\mbox{Log}_P(P_i)=S_i=P^{\frac{1}{2}}\log(P^{-\frac{1}{2}}P_iP^{-\frac{1}{2}})P^{\frac{1}{2}}.
\end{align}

TS mapping converts each 2D EEG trial into a 1D feature vector, so that many machine learning algorithms can be used.

\subsection{Euclidean Alignment (EA)}\label{sect:EA}

EA \cite{he2019transfer,he2019channel} is a state-of-the-art DA approach for BCIs, which reduces the individual differences by aligning the EEG covariance matrices.

Some DA approaches \cite{long2013transfer,drwuTNSRE2016} first find a proper discrepancy measure between different distributions, then learn a shared subspace where the distribution discrepancy is explicitly minimized. Maximum mean discrepancy \cite{gretton2007kernel} is a popular distribution discrepancy measure, which is defined as the distance between the mean feature embeddings of different distributions.

Similar to these maximum mean discrepancy based DA approaches, EA views the covariance matrices as the feature embeddings of different EEG trials, and finds projections to minimize the distance between the mean covariance matrices of different subjects.

For a subject with $N$ trials $\{X_i\}_{i=1}^N$ (each row of $X_i$ is an EEG channel), EA first computes the individual covariance matrices
\begin{align}
  C_i=X_iX_i^T, \quad i=1,2,...,N \label{eq:cov}
\end{align}
and the mean covariance matrix
\begin{align}
  \bar{C}=\frac{1}{N}\sum_{i=1}^{N}C_i.\label{eq:mcov}
\end{align}
The projection matrix for the subject is then
\begin{align}
R=\bar{C}^{-1/2}. \label{eq:ref}
\end{align}
Finally, EA performs the following projection for each trial:
\begin{align}
  \tilde{X}_i=RX_i, \quad i=1, 2, \ldots, N. \label{eq:EA}
\end{align}

After EA, the mean covariance matrix of the subject becomes an identity matrix:
\begin{align}
\frac{1}{N}\sum_{i=1}^{N}\tilde{X}_i\tilde{X}_i^T  &=\frac{1}{N}\sum_{i=1}^{N}RX_iX_i^TR\nonumber \\
&=\bar{C}^{-1/2}\left(\frac{1}{N}\sum_{i=1}^{N}X_iX_i^T\right)\bar{C}^{-1/2}\nonumber\\
&=\bar{C}^{-1/2}\bar{C}\bar{C}^{-1/2}=I. \label{eq:meanC}
\end{align}

After performing EA for all subjects, they share the same mean covariance matrix, i.e., the distances between the mean covariance matrices of different subjects are minimized (they become zero), and hence data distributions from different subjects become more similar.

We can also understand EA as a correction of data shift. If we view each EEG covariance matrix as a point on a Riemannian manifold, then individual differences cause shifts of these points, although they may entail more than just a simple displacement \cite{Zanini2018}. In order to correct this shift, EA moves the covariance matrices of each subject to center them at the identity matrix. The \emph{congruence invariance} property makes sure that the distances among the within-subject covariance matrices remain unchanged. So, EA makes the data distributions from different subjects closer, while preserving the local distance information of each subject.

\section{Label Alignment (LA) for Different Set DA} \label{sect:LA}

This section introduces our proposed LA for different set DA, and discusses its relationship with EA and CORAL \cite{sun2016return}.

\subsection{LA}

Generally, there are three types of data shift in transfer learning:
\begin{enumerate}
\item \emph{Covariate shift} \cite{shimodaira2000improving,sugiyama2007direct}: the distributions of the inputs (features) are different.
\item \emph{Prior probability shift}: the distributions of the output are different.
\item \emph{Concept shift} \cite{utgoff1986shift}: the relationships between the inputs and the output are different.
\end{enumerate}

EA considers only the covariate shift but ignores the other two. Although it has been shown to significantly improve the cross-subject classification performance in \cite{he2019transfer}, it only aligns the data in the feature space, and may not work well when the source subjects and the target subject have different label spaces.

This section proposes LA, which extends EA to different label spaces, by simultaneously considering multiple types of data shift. Its main idea is to independently move the per-class covariance matrices of each source subject, to re-center them at the corresponding class center of the target subject.

More specifically, for an $M$-class classification problem, we assume the source and target subjects have the same number of classes, but their class labels are partially or completely different. Our goal is to use the source data to help the classification of the target trials. LA seeks a transformation matrix $A_m$ for the trials of the $m$-th class ($m=1,2, \cdots, M$) from the source subject, such that the distance between the mean covariance matrices of the corresponding class in different domains are minimized:
\begin{align}
A_m&=\arg\min_A||A\bar{C}_{S,m}A^T-\bar{C}_{T,m}||_F^2, \quad m=1, 2, \cdots, M. \label{eq:Aki}
\end{align}
where $\bar{C}_{S,m}$ is the mean covariance matrix of the $m$-th class of the source subject, and $\bar{C}_{T,m}$ the mean covariance matrix of the $m$-th class of the target subject. In this paper, we use the \emph{Log-Euclidean mean} \cite{arsigny2007geometric}, which is frequently used for symmetric positive definite matrices and much easier to compute than the Riemannian mean.

We adopt the optimization approach in \cite{sun2016return} to solve for $A_m$ in (\ref{eq:Aki}):
\begin{align}
A_m&=\bar{C}_{T,m}^{\frac{1}{2}}\bar{C}_{S,m}^{-\frac{1}{2}}, \quad m=1, 2, \cdots, M. \label{eq:Aki2}
\end{align}

Then, each trial $X_j$ of the source subject is transformed to:
\begin{align}
  \tilde{X}_j=A_mX_j, \quad \mbox{if } X_j\in \mbox{Class } m \label{eq:LA}
\end{align}

The difference between the mean covariance matrices of the corresponding class between the transformed source subject and the target subject becomes
\begin{align}
A\bar{C}_{S,m}A^T-\bar{C}_{T,m}&=
\bar{C}_{T,m}^{\frac{1}{2}}\bar{C}_{S,m}^{-\frac{1}{2}}\bar{C}_{S,m}\bar{C}_{S,m}^{-\frac{1}{2}}\bar{C}_{T,m}^{\frac{1}{2}}-\bar{C}_{T,m}\nonumber \\
&=\bar{C}_{T,m}^{\frac{1}{2}}I\bar{C}_{T,m}^{\frac{1}{2}}-\bar{C}_{T,m}=\mathbf{0}, \label{eq:covdiff}
\end{align}
where $\mathbf{0}$ is an all-zero matrix, i.e., the objective function in (\ref{eq:Aki}) is minimized.

A key question in LA is how to obtain $\bar{C}_{T,m}$, which requires some labeled target domain samples. We consider the following offline classification scenario: we have access to the unlabeled EEG trials (the same assumption is also used in EA), and we can label a few of them to estimate $\bar{C}_{T,m}$. To have a good estimate of $\bar{C}_{T,m}$ from only a few labeled trials, we need to select these trials very carefully. In this paper, we perform $k$-medoids clustering based on the Riemannian distances among the target EEG trials, label the $k$ cluster centers, and then estimate $\bar{C}_{T,m}$ from them. In the rare case that the $k$ centers had fewer than $M$ different labels, we use EA to replace LA.

Another question is how we match the source labels with the target labels. When the source and target label sets partially overlap, for the labels in common, we match each source label with the same target label, and then randomly match each remaining source label with a remaining target label. For example, if the source label set is $\{A,B,C\}$ and the target label set is $\{A,D,E\}$, then we match source label $A$ with target label $A$, source label $B$ with target label $D$ (or $E$), and source label $C$ with target label $E$ (or $D$). If the source and target label sets are completely different, we randomly match the source and target labels.

The pseudo-code of LA is shown in Algorithm~\ref{algo:LA}. We perform LA for each source subject separately if there are multiple source subjects. After LA, the source domain and the target domain have the same label set, and the trials in the same class are aligned. Then, trials from the two domains can be combined directly for feature extraction and classification. Or, an additional DA approach can be applied after LA to further improve the transfer learning performance, as shown in Section~\ref{sect:experiments}.

\begin{algorithm}[htpb]
\KwIn{$\{X_j, y_j\}_{j=1}^{N_S}$, labeled source domain trials\;
\hspace*{10mm} $\{y_{S,m}\}_{m=1}^M$, label set of the source domain\;
\hspace*{10mm} $\{X_i\}_{i=1}^{N_T}$, unlabeled target domain trials\;
\hspace*{10mm} $\{y_{T,m}\}_{m=1}^M$, label set of the target domain\;
\hspace*{10mm} $k$, number of target domain trials to be labeled.}
\KwOut{$\{\widetilde{X}_j, \widetilde{y}_j\}_{j=1}^{N_S}$, aligned source domain trials.}
Compute the target domain covariance matrices $\{C_i\}_{i=1}^{N_T}$ by (\ref{eq:cov})\;
Perform $k$-medoids clustering on $\{C_i\}_{i=1}^{N_T}$ using the Riemannian distance\;
Label the $k$ medoids\;
Compute $\bar{C}_{T,m}$ ($m=1, 2, \ldots, M$), the mean covariance matrix of each target domain class from the $k$ labeled medoids\;
Compute $\bar{C}_{S,m}$ ($m=1, 2, \cdots, M$), the mean covariance matrix of each source domain class\;
Match each source domain label with a target domain label; assume $y_{S, m}$ is matched with $y_{T,m}$\;
\For{$m=1, 2, \ldots, M$}
{Compute $A_m$ by (\ref{eq:Aki2})\;
 Compute $\{\widetilde{X}_j\}_{j=1}^{N_S}$ by (\ref{eq:LA}) and set $\widetilde{y}_j=y_{T,m}$, $\forall_j y_j=y_{S,m}$\;
}
\textbf{Return} $\{\widetilde{X}_j, \widetilde{y}_j\}_{j=1}^{N_S}$
\caption{LA for different set domain adaptation.}
\label{algo:LA}
\end{algorithm}

\subsection{LA versus EA}

The difference between LA and EA is illustrated in Fig.~\ref{fig:LAdiagram}. For clarity, binary classification is used, but both EA and LA can be easily extended to multi-class classification, as shown later in this paper.
In Fig.~\ref{fig:LAdiagram}, each EEG trial is represented by its covariance matrix, as a point on a Riemannian manifold. The source domain (blue points) and target domain (black points) represent two different subjects, who have trials from different motor imagery tasks (indicated by different shapes of the points. Note that the shapes in the target domain are only used to help understand our approach, but not to suggest that we need to know all target labels). Initially, the source and target domains scatter far away from each other, due to the domain gap and also the category gap. If we build a classifier on the source domain (indicated by the red dashed line) and apply it directly to the target domain, it may not work at all. EA and LA alleviate this problem by reducing the gaps between the two domains before classification:

\begin{figure}[htpb]\centering
\includegraphics[width=\linewidth,clip]{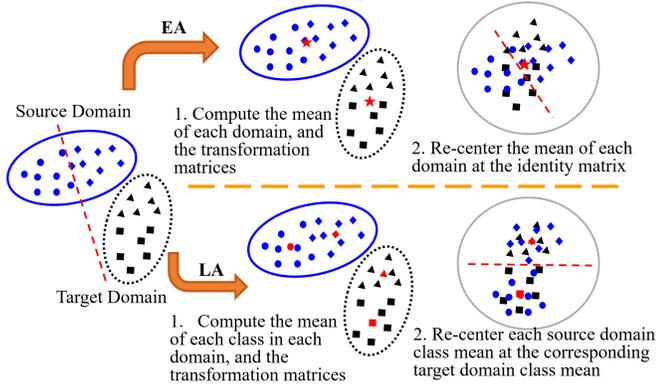}
\caption{Illustration of EA and LA. } \label{fig:LAdiagram}
\end{figure}

\begin{enumerate}
  \item EA focuses on the domain gap but ignores the category gap completely. It first computes the mean covariance matrix of each domain (indicated by the red stars), from which a transformation matrix of each domain is computed. Using the transformation matrix, EA then re-centers each domain at the identity matrix, and makes the source and target domains overlap with each other, i.e., the domain gap between them is reduced. If we build a classifier in the source domain (the red dashed line) and apply it to the target domain, the classification performance would be improved.
  \item LA considers the domain gap and the category gap simultaneously. It first computes the mean covariance matrix of each source domain class (indicated by the red circle and the red diamond), and estimates the mean covariance matrix of each target domain class (indicated by the red triangle and the red square). Then, LA re-centers each source domain class at the corresponding estimated class mean of the target domain. If we build a classifier in the source domain (the red dashed line) and apply it to the target domain, the classification performance would be further improved.
\end{enumerate}

\subsection{LA versus CORrelation ALignment (CORAL)}

Sun \emph{et al.} \cite{sun2016return} proposed an unsupervised DA approach, CORrelation ALignment (CORAL), to minimize the domain shift by aligning the second-order statistics of different distributions.

Given a source domain $D_S\in\mathbb{R}^{N_S\times d}$ and a target domain $D_T\in\mathbb{R}^{N_T\times d}$, where $N_S$ and $N_T$ are the number of trials in the source domain and the target domain, respectively, and $d$ the feature dimensionality. CORAL first computes the feature covariance matrix $C_S\in\mathbb{R}^{d\times d}$ in the source domain and $C_T\in\mathbb{R}^{d\times d}$ in the target domain. Then, it finds a linear transformation matrix $A\in\mathbb{R}^{d\times d}$ for the source domain features, so that the Frobenius norm of the difference between the covariance matrices of the two domains is minimized, i.e.,
\begin{align}
\min_A||A^TC_SA-C_T||_F^2 \label{eq:CORAL}
\end{align}

Although (\ref{eq:CORAL}) seems similar to the objective function of LA in (\ref{eq:Aki}), they are different:
\begin{enumerate}
  \item CORAL uses 1D features, and each domain has only one feature covariance matrix, which measures the covariances between different pairs of individual features. LA uses 2D features (EEG trials), and each EEG trial has a covariance matrix, which measures the covariances between different pairs of EEG channels. So, the covariance matrices in CORAL and LA have different meanings.
  \item CORAL minimizes the distance between the covariance matrices in different domains, whereas LA minimizes the distance between the mean covariance matrices of the corresponding class in different domains.
  \item CORAL works when the source domain has the same class labels as the target domain, and it finds one transformation matrix for each source domain. LA considers the case that the source and target domains have different class labels (of course, it also works when the two domains have the same class labels), and it finds one transformation matrix for each class of the source domain.
\end{enumerate}

In summary, LA and CORAL have different inputs, different optimization objectives, and also different application scenarios. When the source and target domains have the same class labels, each 2D EEG trial can be mapped from the Riemannian manifold to the tangent space to obtain a 1D feature vector, and hence be plugged into CORAL. However, CORAL cannot be used when the source and target domains have different labels.

\section{Datasets} \label{sect:datasets}

This section describes and visualizes the two motor imagery datasets used in our experiments.

\subsection{Datasets and Preprocessing} \label{sect:Preprocessing}

\begin{figure}[htpb]\centering
\includegraphics[width=\linewidth,clip]{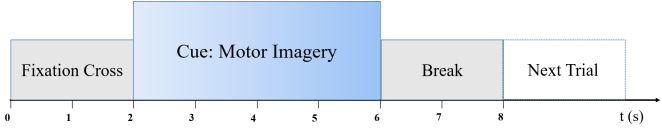}
\caption{Timing scheme of the motor imagery tasks.} \label{fig:paradigm}
\end{figure}

Both datasets were from BCI Competition IV\footnote{http://www.bbci.de/competition/iv/.}, and were collected in a cue-based setting. In each experiment, the subject was sitting in front of a computer and performed motor imagery tasks at the prompt of visual cues, as shown in Fig.~\ref{fig:paradigm}. Each trial began when a fixation cross appeared on the black screen ($t=0$), which prompted the subject to be prepared. After a short period, an arrow pointing to a certain direction was displayed as the visual cue ($t=2$). The cue was displayed for a few seconds, during which the subject was instructed to perform the desired motor imagery task according to the direction of the arrow. The subject stopped the motor imagery when the visual cue disappeared ($t=6$). A short break followed, until the next trial began ($t=8$).

%In classification, a trial only contained the segment of EEG signal that a subject was performing a motor imagery task, and its label was the corresponding motor imagery task, e.g., ``left hand" or ``right hand".

The first dataset\footnote{http://www.bbci.de/competition/iv/desc\_1.html.} (Dataset~1 \cite{Blankertz2007}) was recorded from seven healthy subjects by 59 EEG channels at 100 Hz. Each subject was instructed to perform two classes of motor imagery tasks, which were selected from three options: left hand, right hand, and feet. The recording of each subject was divided into three sessions: calibration, evaluation, and special feature. This paper only used the calibration data, because they included complete label information. Each subject had 100 trials from each class.

The second dataset\footnote{http://www.bbci.de/competition/iv/desc\_2a.pdf.} (Dataset~2a) was recorded from nine healthy subjects by 22 EEG channels and 3 EOG channels at 250 Hz (we downsampled it to 100 Hz, to be consistent with Dataset~1). Each subject was instructed to perform four classes of motor imagery tasks: left hand, right hand, both feet, and tongue, which were represented by labels 1, 2, 3 and 4, respectively. A training session and an evaluation session were recorded on different days for each subject. We only used the 22-channel EEG data in the training session, which included complete label information. Each subject had 72 trials from each class, and 288 trials in total.

For both datasets, the EEG signals were preprocessed using the Matlab EEGLAB toolbox \cite{Delorme2004}, following the guideline in \cite{Blankertz2008}. First, a causal band-pass filter (20-order linear phase Hamming window FIR filter designed by Matlab function fir1, with 6dB cut-off frequencies at [8, 30] Hz) was applied to remove muscle artifacts, line-noise contamination and DC drift. Next, we extracted EEG signals between $[0.5, 3.5]$ seconds after the cue appearance as our trials.

Table~\ref{tab:datasets} summaries the characteristics of the two datasets.

\begin{table}[htpb] \centering \setlength{\tabcolsep}{1.5mm}
\caption{summary of the two motor imagery datasets.}   \label{tab:datasets}
\begin{tabular}{c|ccccc}   \hline
& \multicolumn{5}{|c}{Number of} \\ \cline{2-6}
& Channels & Time samples & Subjects & Classes & Trials/class \\ \hline
Data 1 & 59&300 &7 &2&100\\ \hline
Data 2a & 22 & 300&9 &4&72\\ \hline
\end{tabular}
\end{table}

\subsection{Data Visualization} \label{sect:visualization}

In order to intuitively show how EA and LA reduce the distribution discrepancies between the target and source subjects, we first projected the EEG covariance matrices from the Riemannian manifold into the tangent space, then used the 1D tangent vectors as features to represent the EEG trials, as introduced in Section~\ref{sect:TS}. Finally, we used $t$-stochastic neighbor embedding ($t$-SNE) \cite{Maaten2008}, a technique for dimensionality reduction and high-dimensional dataset visualization, to display the EEG trials (tangent vectors) before and after EA/LA in 2D.

\begin{figure}[htpb]\centering
\includegraphics[width=\linewidth,clip]{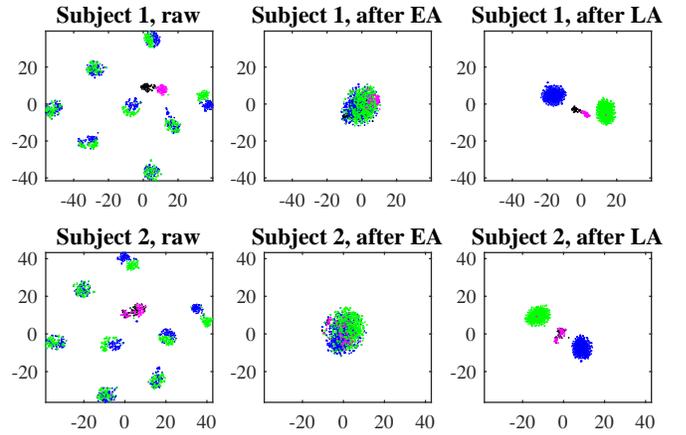}
\caption{$t$-SNE visualization when the first two subjects were used as the target subjects, respectively. Red dots: trials of Label~2 from the target subject; black dots: trials of Label~3 from the target subject; blue dots: trials of Label~1 from the source subjects; green dots: trials of Label~4 from the source subjects. The first column shows the trials without alignment, the second column shows the trials after EA, and the third after LA.} \label{fig:tSNE}
\end{figure}

More specifically, we first divided Dataset~2a into two datasets with different label spaces: the source dataset consisted of trials with Labels~1 and 4, and the target dataset with Labels~2 and 3. Then, we picked one subject from the target dataset as the target subject, and the remaining eight subjects from the source dataset as the source subjects. Fig.~\ref{fig:tSNE} shows two examples when the first two subjects were used as the target subjects, respectively. The red and black dots are trials of Labels~2 and 3 from the target subject, respectively. The blue and green dots are trials of Labels~1 and 4 from the source subjects, respectively. The first column shows the trials without alignment, the second column shows the trials after EA, and the third after LA.

Observe that trials from the source subjects (blue and green dots) are scattered far away from those of the target subject (red and black dots), when no alignment is performed. However, the target and source trials overlap with each other after EA, since their centers are now identical. After LA, the target and source trials are further aligned according to their labels. It's clear that different classes are more distinguishable after LA.

\section{Experiments and Results} \label{sect:experiments}

This section presents performance comparisons of LA with other approaches on the two datasets. The code is available at \url{https://github.com/hehe91/LA}.

\subsection{Domain Adaptation (DA) Scenarios} \label{sect:protocol}

We investigated the problem that the source and target subjects have different label spaces, and considered the following five DA scenarios:
\begin{enumerate}
  \item \emph{Scenario I-a}: The source and target subjects have the same feature space and partially overlapping label spaces (binary classification).
  \item \emph{Scenario I-b}: The source and target subjects have the same feature space and partially overlapping label spaces (multi-class classification).
  \item \emph{Scenario II-a}: The source and target subjects have the same feature space and completely different label spaces (binary classification).
  \item \emph{Scenario II-b}: The source and target subjects have the same feature space and completely different label spaces (multi-class classification).
  \item \emph{Scenario III}: The source and target subjects have different feature spaces and also different label spaces.
\end{enumerate}

For Scenarios~I-a, I-b, II-a and II-b, in each experiment we divided Dataset~2a into two sub-datasets, a source dataset and a target dataset, such that they had the same feature space and different label spaces. Each sub-dataset was named by its label space, for example, sub-dataset ``1, 2" consisted of trials with Labels 1 and 2 only, and sub-dataset ``3, 4" consisted of trials with Labels 3 and 4 only. Then, ``1, 2$\rightarrow$3, 4" denotes the experiment that Sub-dataset ``1, 2" was used as the source dataset and Sub-dataset ``3, 4" the target dataset.

Then, the datasets used in the five DA scenarios were:
\begin{enumerate}
  \item \emph{Scenario I-a}: We divided Dataset~2a into a source sub-dataset and a target sub-dataset, ensuring they had one identical label and one different label. There were 24 such sub-dataset combinations in total, e.g., ``1, 2$\rightarrow$1, 3" and ``1, 2$\rightarrow$3, 2".
  \item \emph{Scenario I-b}: We divided Dataset~2a into a source sub-dataset and a target sub-dataset, ensuring they had two identical labels and a different label. There were 12 such combinations in total, e.g., ``1, 2, 3$\rightarrow$1, 2, 4" and ``1, 2, 3$\rightarrow$1, 4, 3".
  \item \emph{Scenario II-a}: We divided Dataset~2a into a source sub-dataset and a target sub-dataset, ensuring they had completely different labels. There were six such combinations in total, e.g., ``1, 2$\rightarrow$3, 4" and ``2, 3$\rightarrow$1, 4".
  \item \emph{Scenario II-b}: We used the same sub-dataset combinations as in Scenario~I-b, but mismatched the labels between the target and source subjects, e.g., ``1, 2, 3$\rightarrow$2, 1, 4" and ``1, 2, 4$\rightarrow$2, 1, 3".
  \item \emph{Scenario III}: We used Dataset 1 as the source dataset and sub-dataset ``3, 4" of Dataset 2a as the target dataset, so that they had different feature spaces (their EEG channels were different) and also different label spaces.
\end{enumerate}

Once a dataset choice was made, each time we picked one subject from the target dataset as the target subject, and the remaining subjects from the source dataset as the source subjects. As the target dataset always had nine subjects, we had nine sub-experiments for each dataset combinations. Table~\ref{tab:protocol} summaries the characteristics of all scenarios, where $k$ is the number of labeled target subject trials.

\begin{table}[htpb] \centering \setlength{\tabcolsep}{1.5mm}
\caption{Summary of the four DA scenarios, where $k$ is the number of labeled target subject trials.}   \label{tab:protocol}
\begin{tabular}{l|ccccc}   \hline
&No. of dataset & No. of sub- & No. of  & No. of  \\
&  combinations&experiments &training trials&test trials\\\hline
Scenario I-a & $24$&$24\times9$ &$144\times8+k$ &$144-k$\\
Scenario I-b & $12$ & $12\times9$&$216\times8+k$ &$216-k$\\
Scenario II-a & $6$&$6\times9$ &$144\times8+k$ &$144-k$\\
Scenario II-b & $12$ & $12\times9$&$216\times8+k$ &$216-k$\\
Scenario III & $1$ & $9$&$1400+k$ &$144-k$\\ \hline
\end{tabular}
\end{table}

\subsection{Experimental Settings}\label{sect:settings}

We first divided the BCI classification pipeline into three stages:
\begin{enumerate}
  \item \emph{Preprocessing}, which first temporally filters the EEG data, then epochs the continuous EEG signals into trials, as described in Section~\ref{sect:Preprocessing}.
  \item \emph{Alignment}, which selectively performs different alignments.
  \item \emph{Classification}, which extracts features and trains classifiers.
\end{enumerate}

In order to emphasize the effect of LA, the algorithms to be compared consisted of the same preprocessing and classification stages, but different alignments. More specifically, three alignment approaches were compared:
\begin{enumerate}
  \item \emph{Raw}, which did not perform any alignment.
  \item \emph{EA}, which performed EA.
  \item \emph{LA}, which performed LA.
\end{enumerate}

In each scenario, the experiments were designed to answer the following two questions:
\begin{enumerate}
  \item[] \emph{Question 1: Can LA be used as an effective preprocessing step before different feature extraction and classification algorithms?}
  \item[] \emph{Question 2: Can LA be integrated with other DA approaches to further improve the classification performance?}
\end{enumerate}

For Question 1, we used two feature extraction and classification pipelines:
\begin{enumerate}
  \item CSP-LDA: It spatially filtered the EEG data by CSP, computed the log-variance as features, and then used them in an LDA classifier.
  \item TS-SVM: It extracted the Riemannian TS features, as introduced in Section~\ref{sect:TS}, then used them in an SVM classifier.
\end{enumerate}

Combining these two pipelines with the three alignment approaches (Raw, EA, LA), we had $2\times3=6$ algorithms to be compared. Our goal was to verify whether LA performs the best in both pipelines.

For Question 2, we first extracted the Riemannian TS features, and then used different DA approaches in classification stage (because they need 1D features):
\begin{enumerate}
  \item BL (baseline), which directly applied an SVM classifier to the TS features, without any additional DA approach.
  \item JDA, which applied JDA to the TS features, and then used them in an SVM classifier.
  \item JGSA, which applied JGSA to the TS features, and then used them in an SVM classifier.
  \item MEDA, which applied MEDA to the TS features.
\end{enumerate}

Combining these four approaches with the three alignments (Raw, EA, LA), we had $4\times3=12$ algorithms to be compared. Our goal was to verify whether ``LA+JDA/JGSA/MEDA\textgreater LA+BL\textgreater Raw+JDA/JGSA/MEDA", where ``\textgreater" means ``\textit{outperform}". For example, ``LA+BL\textgreater Raw+JDA/JGSA/MEDA" means LA outperforms classical DA approaches such as JDA, JGSA and MEDA, and ``LA+JDA/JGSA/MEDA\textgreater LA+BL" means the performance could be further improved by integrating LA with other DA approaches, i.e., LA is compatible with and complementary to other DA approaches.

\subsection{Scenario~I-a: Same Feature Space and Partially Overlapping Label Spaces in Binary Classification} \label{sect:scenario1-a}

This subsection considers the binary classification scenario that the source and target subjects have the same feature space and partially overlapping label spaces. As introduce in Section~\ref{sect:protocol}, we had 24 sub-dataset combinations to be tested.

Because in Scenario~I-a the source subjects had one identical label and one different label from the target subject, we first matched the identical label, then the remaining labels. For example, in the combination ``1, 2 $\rightarrow$ 1, 3", we matched source Label~1 with target Label~1, and source Label~2 with target Label~3. For algorithms without LA, we directly assigned Label~3 to the source trials with Label~2. For algorithms with LA, we first aligned the source Label~1 trials with the target Label~1 trials, then aligned the source Label~2 trials with the target Label~3 trials, and assigned Label~3 to the source trials of Label~2.

For algorithms involving LA, we considered $k\in\{2,4,...,20\}$ in $k$-medoids clustering of LA in Section~\ref{sect:LA}. In the rare case that the labeled target trials came from the same class, we cannot perform LA as there was not enough information to estimate the two class means of the target subject; thus, we performed EA instead of LA. No matter whether the labeled target trials were used in the alignment or not, they were always combined with the labeled source trials for feature extraction and classification. All labeled target subject trials were excluded from the test set, so all algorithms had the same training set and test set.

\emph{\textbf{Question 1:} Can LA be used as an effective preprocessing step before different feature extraction and classification algorithms?}

\begin{figure*}[htpb]\centering
\includegraphics[width=\linewidth,clip]{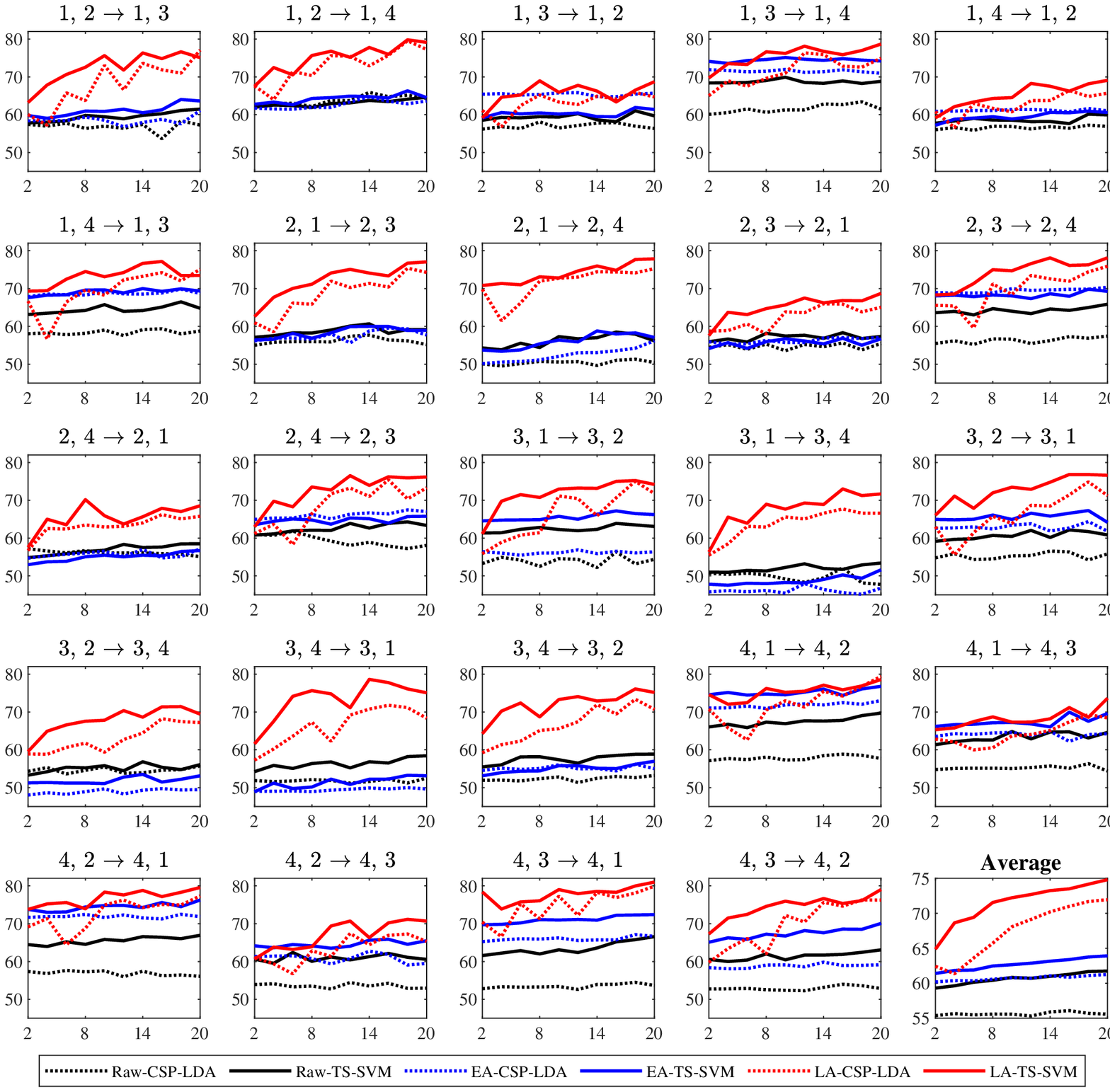}
\caption{Average classification accuracies (\%) in Scenario~I-a. The horizontal axis indicates the number of labeled target subject trials ($k$ in $k$-medoids clustering in Section~\ref{sect:LA}), and the vertical axis the classification accuracies.} \label{fig:1-a}
\end{figure*}

We compared Raw, EA, and LA in the two classification pipelines to answer this question. Fig.~\ref{fig:1-a} shows the performances of the six algorithms on the 24 different sub-dataset combinations, where each subfigure shows the average accuracies across the nine subjects (each as the target subject once). The last subfigure shows the average performances across the 24 experiments. Observe that:
\begin{enumerate}
  \item EA-CSP-LDA outperformed Raw-CSP-LDA on 20 out of the 24 experiments, and EA-TS-SVM outperformed Raw-TS-SVM on 14 out of the 24 experiments. On average EA-CSP-LDA outperformed Raw-CSP-LDA, and EA-TS-SVM outperformed Raw-TS-SVM. This suggests EA was generally effective, but not always, when the source and target label spaces were different.
  \item When $k$ became large, LA-CSP-LDA outperformed Raw-CSP-LDA in all 24 experiments, and LA-TS-SVM also outperformed Raw-TS-SVM in all 24 experiments. These suggest that LA was able to cope well with partially different label spaces.
  \item When $k$ became large, LA-CSP-LDA outperformed EA-CSP-LDA on all 24 experiments, and LA-TS-SVM also outperformed EA-TS-SVM on all 24 experiments. This suggests LA was more effective and robust than EA.
  \item Generally, the classification accuracies of LA-CSP-LDA and LA-TS-SVM increased when there were more labeled target trials for estimating the class means, which is intuitive.
\end{enumerate}

We also performed statistical tests to determine if the differences between the LA-based algorithms and others were statistically significant. We first defined an aggregated performance measure called the area under the curve (AUC). For a particular algorithm on a particular subject, the AUC was the area under its accuracy curve when the number of labeled target subject trials increased from 2 to 20. Subjects from all 24 experiments were concatenated, so we had $24\times9=216$ subjects in total. Each algorithm had 216 AUCs. We then performed paired $t$-tests on these AUCs. The null hypothesis was that the difference between the paired samples has zero mean, which was rejected if $p < \alpha$, where $\alpha = 0.05$ was used. The results are shown in Table~\ref{tab:ttest1-a}, where the statistically significant ones are marked in bold. LA-CSP-LDA significantly outperformed EA-CSP-LDA, and LA-TS-SVM significantly outperformed EA-TS-SVM. These results echo the observations from Fig.~\ref{fig:1-a} and answer Question~1: LA can be used as an effective preprocessing step before different feature extraction and classification algorithms.

\begin{table}[htpb] \centering \setlength{\tabcolsep}{2.5mm}
\caption{Scenario~I-a: $p$-values of paired $t$-tests on the AUCs of the classification accuracy curves in Fig.~\ref{fig:1-a}. The null hypothesis was rejected if $p< \alpha$, where $\alpha=0.05$.}   \label{tab:ttest1-a}
\begin{tabular}{c|cc}   \hline
 & EA-CSP-LDA & EA-TS -SVM \\
\hline
 LA-CSP-LDA & \textbf{0.0000}&\\ \hline
 LA-TS-SVM & & \textbf{0.0000} \\ \hline
\end{tabular}
\end{table}

\emph{\textbf{Question 2: }Can LA be integrated with other DA approaches to further improve the clssification performance?}

\begin{figure*}[htpb]\centering
\includegraphics[width=\linewidth,clip]{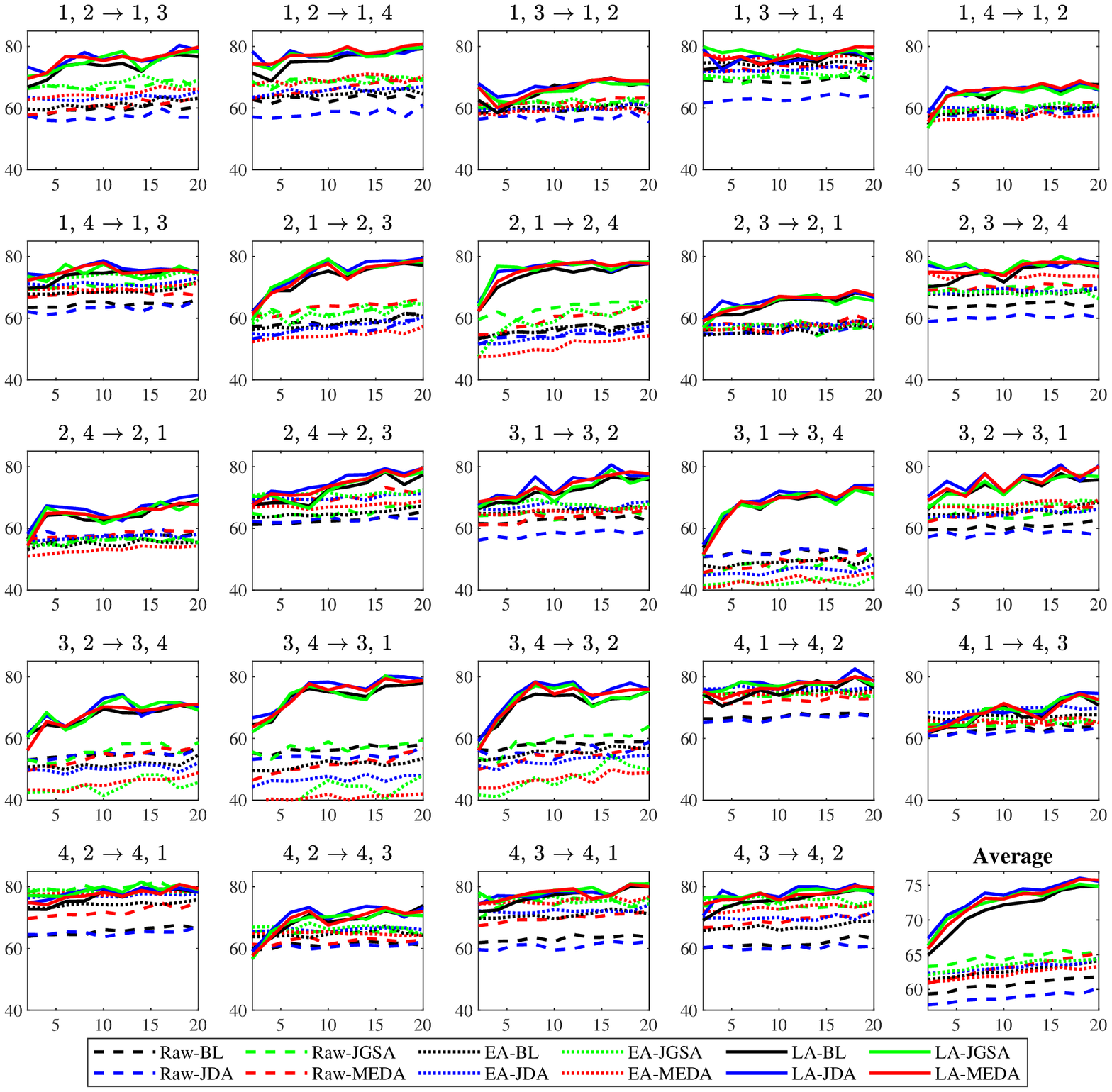}
\caption{Average classification accuracies (\%) in Scenario~I-a, when additional DA approaches were used after LA. The horizontal axis indicates the number of labeled target subject trials ($k$ in $k$-medoids clustering in Section~\ref{sect:LA}), and the vertical axis the classification accuracies.} \label{fig:1TL-a}
\end{figure*}

As introduced in previous subsection, we had 12 algorithms to be compared. We used the same target and source subjects as introduced in Question~1, which resulted in 24 experiments again. Fig.~\ref{fig:1TL-a} shows the performances of the 12 algorithms in the 24 experiments, where each subfigure shows the average accuracies across the nine subjects, and the last subfigure shows the average performances across the 24 experiments.

Observe that:
\begin{enumerate}
  \item When $k$ was large, LA-BL always outperformed Raw-BL and EA-BL, LA-JDA always outperformed Raw-JDA and EA-JDA, LA-JGSA always outperformed Raw-JGSA and EA-JGSA, and LA-MEDA always outperformed Raw-MEDA and EA-MEDA. These suggest that LA was effective regardless of whether additional DA approaches were used or not.
  \item LA-BL always outperformed Raw-JDA and Raw-MEDA, and outperformed Raw-JGSA in 23 out of 24 experiments. These suggest that LA can outperform classical DA approaches such as JDA, JGSA and MEDA.
  \item Both LA-JDA and LA-MEDA always outperformed LA-BL, and LA-JGSA outperformed LA-BL in 23 out of the 24 experiments. These suggest that it may be advantageous to integrate other DA approaches with LA.
\end{enumerate}

We also performed paired \emph{t}-tests on the AUCs in Fig. \ref{fig:1TL-a}. The results are shown in Table~\ref{tab:ttest1-a-TL}, which indicate that the algorithms involving LA (i.e., LA-BL, LA-JDA, LA-JGSA, LA-MEDA) significantly outperformed those involving EA (i.e., EA-JDA, EA-JGSA, EA-MEDA), and the algorithms combining LA and additional DA approaches (i.e., LA-JDA, LA-JGSA, LA-MEDA) significantly outperformed LA-BL. These results echo the observations from Fig. \ref{fig:1TL-a} and answer Question~2: LA can not only outperform EA and classical DA approaches, but the classification performance can be further improved when integrated with other DA approaches.

\begin{table}[htpb] \centering \setlength{\tabcolsep}{2.5mm}
\caption{Scenario~I-a: $p$-values of paired $t$-tests on the AUCs of the classification accuracy curves in Fig.~\ref{fig:1TL-a}. The null hypothesis was rejected if $p< \alpha$, where $\alpha=0.05$.}   \label{tab:ttest1-a-TL}
\begin{tabular}{c|cccc}   \hline
 & LA-BL & EA-JDA &EA-JGSA & EA-MEDA  \\ \hline
 LA-BL & &\textbf{0.0000} &\textbf{0.0000} &\textbf{0.0000}\\ \hline
 LA-JDA &\textbf{0.0000} & \textbf{0.0000} & &\\ \hline
 LA-JGSA &\textbf{0.0001}&&\textbf{0.0000}&\\ \hline
 LA-MEDA &\textbf{0.0000}&&&\textbf{0.0000}\\ \hline
\end{tabular}
\end{table}

\subsection{Scenario~I-b: Same Feature Space and Partially Overlapping Label Spaces in Multi-Class Classification} \label{sect:scenario1-b}

This subsection considers the multi-class classification scenario that the source and target subjects have the same feature space and partially overlapping label spaces. As introduced in Section~\ref{sect:protocol}, we had 12 sub-dataset combinations to be tested.

\emph{\textbf{Question 1:} Can LA be used as an effective preprocessing step before different feature extraction and classification algorithms?}

Again, we compared Raw, EA, and LA in the two classification pipelines to answer this question. CSP filtering was extended from binary classification to multi-class classification by the one-versus-the-rest approach \cite{Dornhege2004}. As we had three class centers of the target subject to be estimated in LA, we considered $k\in\{3,6,...,30\}$ in $k$-medoids clustering.

\begin{figure*}[htpb]\centering
\includegraphics[width=\linewidth,clip]{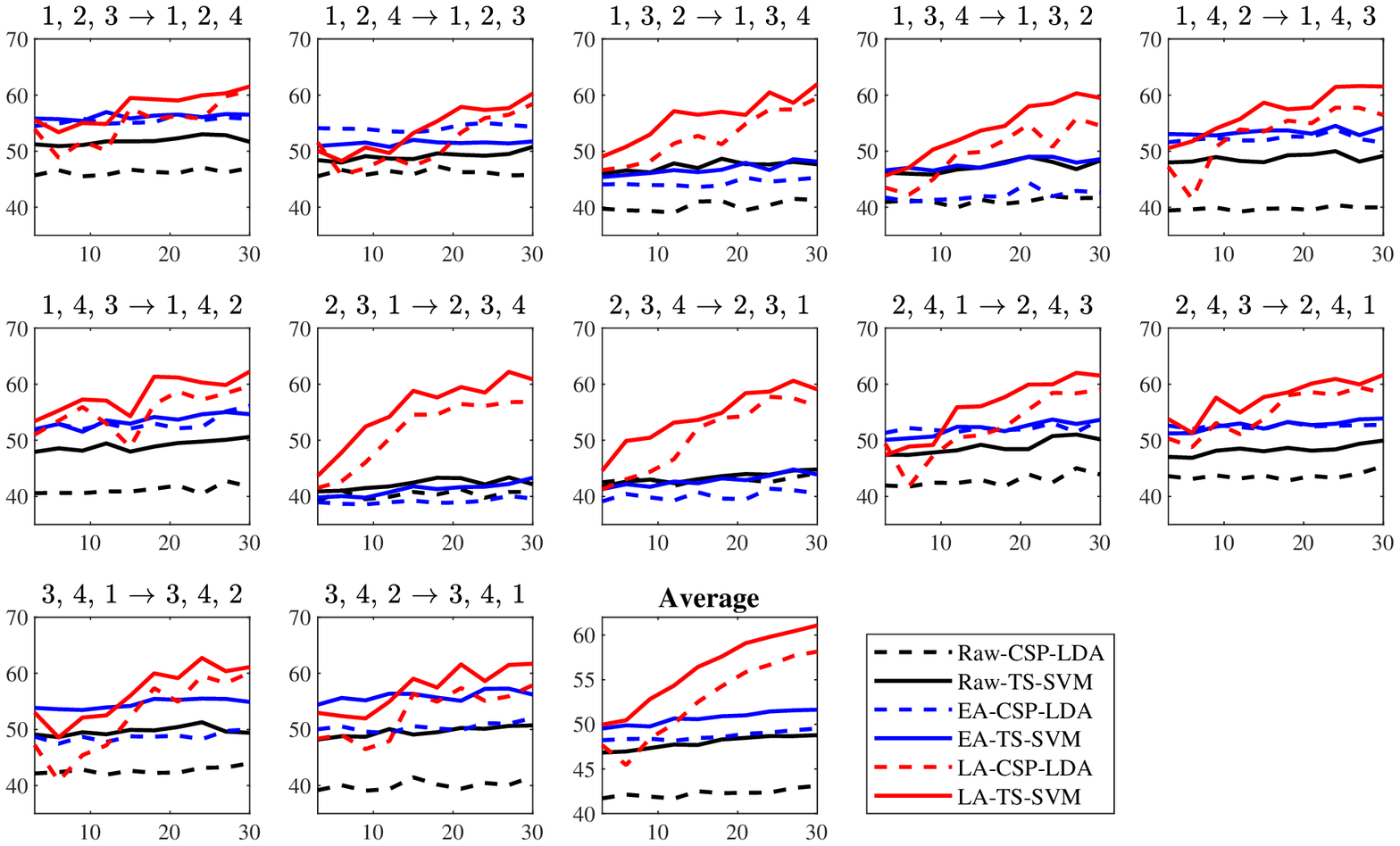}
\caption{Average classification accuracies (\%) in Scenario~I-b. The horizontal axis indicates the number of labeled target subject trials ($k$ in $k$-medoids clustering in Section~\ref{sect:LA}), and the vertical axis the classification accuracies.} \label{fig:1-b}
\end{figure*}

Fig.~\ref{fig:1-b} shows the performances of the six algorithms on the 12 different sub-dataset combinations, where each subfigure shows the average classification accuracies across the nine subjects (each as the target subject once). The last subfigure shows the average performances across the 12 experiments. LA-CSP-LDA always outperformed Raw-CSP-LDA and EA-CSP-LDA, and LA-TS-SVM always outperformed Raw-TS-SVM and EA-TS-SVM. These suggest that LA was effective with different feature extraction and classification algorithms.

Paired \emph{t}-tests on the AUCs in Fig.~\ref{fig:1-b} were also performed to check if the differences between different algorithms were statistically significant. Here the AUC was the area under the accuracy curve when the number of labeled target subject trials increased from 3 to 30. Each algorithm had $12\times9=108$ AUCs. The results are shown in Table~\ref{tab:ttest1-b}, which indicate that LA-CSP-LDA significantly outperformed EA-CSP-LDA, and LA-TS-SVM significantly outperformed EA-TS-SVM.

\begin{table}[htpb] \centering \setlength{\tabcolsep}{2.5mm}
\caption{Scenario~I-b: $p$-values of paired $t$-tests on the AUCs of the classification accuracy curves in Fig.~\ref{fig:1-b}. The null hypothesis was rejected if $p< \alpha$, where $\alpha=0.05$.}   \label{tab:ttest1-b}
\begin{tabular}{c|cc}   \hline
 & EA-CSP-LDA & EA-TS-SVM  \\ \hline
 LA-CSP-LDA & \textbf{0.0196}&\\ \hline
 LA-TS-SVM & & \textbf{0.0010} \\ \hline
\end{tabular}
\end{table}

\emph{\textbf{Question 2:} Can LA be integrated with other DA approaches to further improve the classification performance?}

\begin{figure*}[htpb]\centering
\includegraphics[width=\linewidth,clip]{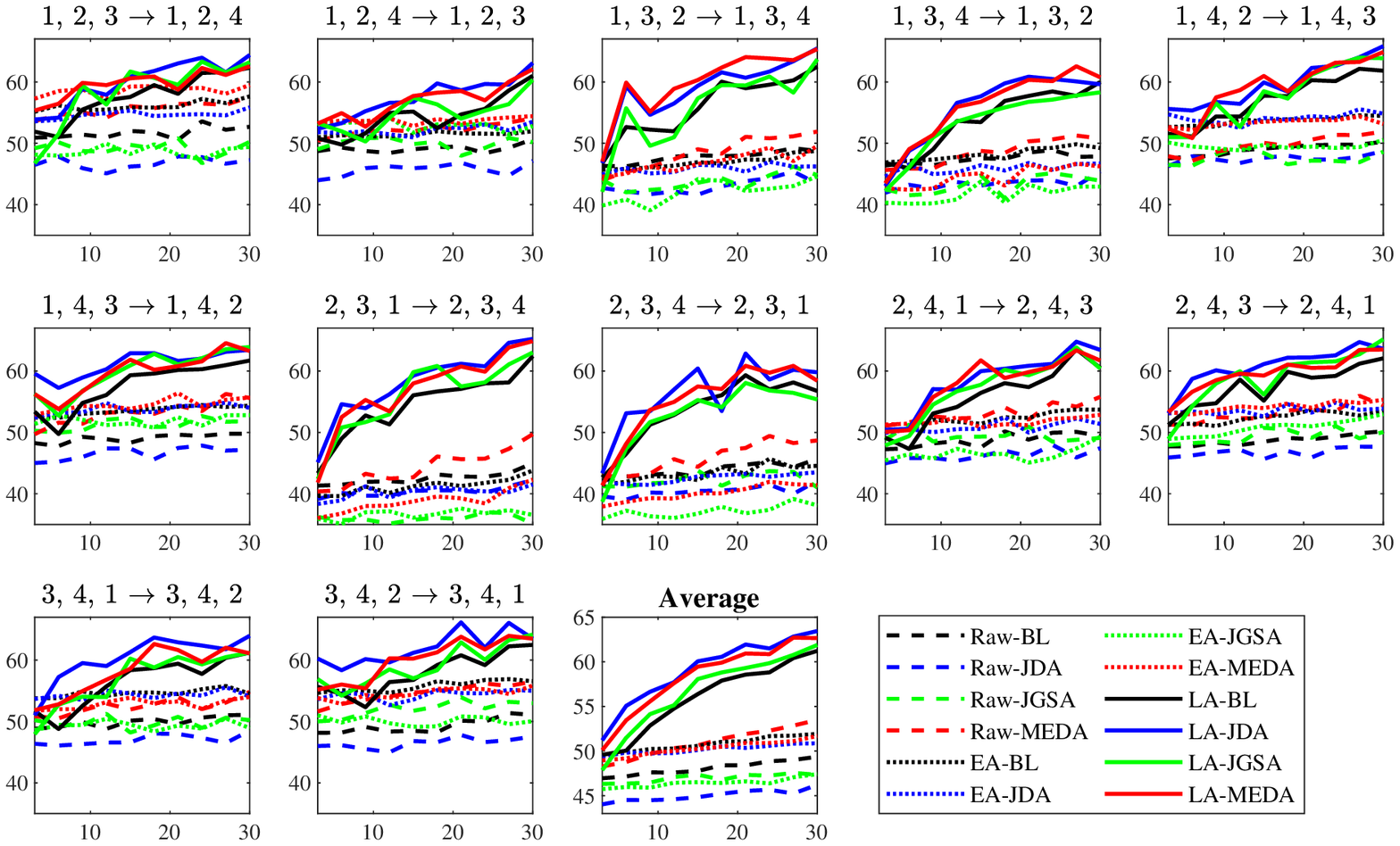}
\caption{Average classification accuracies (\%) in Scenario~I-b, when other DA approaches were used after LA. The horizontal axis indicates the number of labeled target subject trials ($k$ in $k$-medoids clustering in Section~\ref{sect:LA}), and the vertical axis the classification accuracies.} \label{fig:1TL-b}
\end{figure*}

Again, we combined Raw, EA, LA with different DA approaches and obtained 12 algorithms to be compared. Fig.~\ref{fig:1TL-b} shows their performances on the 12 sub-dataset combinations, and the average across the 12 experiments. Observe that:
\begin{enumerate}
  \item When $k$ was large, LA-BL always outperformed Raw-BL and EA-BL, LA-JDA always outperformed Raw-JDA and EA-JDA, LA-JGSA always outperformed Raw-JGSA and EA-JGSA, and LA-MEDA always outperformed Raw-MEDA. These suggest that LA was effective regardless of whether additional DA approaches were used or not.
  \item When $k$ was large, LA-BL outperformed Raw-JDA, Raw-JGSA and Raw-MEDA in all 12 experiments, suggesting that LA can outperform classical DA approaches such as JDA, JGSA and MEDA.
  \item Generally, LA-JDA, LA-JGSA and LA-MEDA outperformed LA-BL, suggesting that it may be advantageous to integrate additional DA approaches with LA.
\end{enumerate}

Table~\ref{tab:ttest1-b-TL} shows the results of paired \emph{t}-tests on the AUCs in Fig. \ref{fig:1TL-b}. The conclusions in binary classification still hold in multi-class classification: LA significantly outperformed EA and classical DA approaches, and its performance can be further significantly improved when integrated with other DA approaches.

\begin{table}[htpb] \centering \setlength{\tabcolsep}{2.5mm}
\caption{Scenario~I-b: $p$-values of paired $t$-tests on the AUCs in Fig.~\ref{fig:1TL-b}. The null hypothesis was rejected if $p< \alpha$, where $\alpha=0.05$.}   \label{tab:ttest1-b-TL}
\begin{tabular}{c|cccc}   \hline
 & LA-BL & EA-JDA &EA-JGSA & EA-MEDA  \\  \hline
 LA-BL & &\textbf{0.0000} &\textbf{0.0000} &\textbf{0.0000}\\ \hline
 LA-JDA &\textbf{0.0000} & \textbf{0.0000} & &\\ \hline
 LA-JGSA &\textbf{0.0295}&&\textbf{0.0000}&\\ \hline
 LA-MEDA &\textbf{0.0000}&&&\textbf{0.0000}\\ \hline
\end{tabular}
\end{table}

\subsection{Scenario~II-a: Same Feature Space and Completely Different Label Spaces in Binary Classification} \label{sect:scenario2}

This subsection considers the scenario that the source and target subjects have the same feature space but completely different label spaces. As introduced in Section~\ref{sect:protocol}, we had six such sub-dataset combinations to be tested. $k\in\{2,4,...,20\}$ in $k$-medoids clustering of LA in binary classification was used.

Because in Scenario~II-a the source subjects had completely different labels from the target subject, the source labels and the target labels were randomly matched for LA. For example, in the experiment ``1, 2$\rightarrow$3, 4", we could align the trials of Label~1 with those of Label~3, and the trials of Label~2 with those of Label~4. We could also align the trials of Label~2 with those of Label~3, and the trials of Label~1 with those of Label~4. Our experiments showed that LA was effective in both alignment strategies.

\emph{\textbf{Question 1:} Can LA be used as an effective preprocessing step before different feature extraction and classification algorithms?}

\begin{figure}[htpb]\centering
\includegraphics[width=\linewidth,clip]{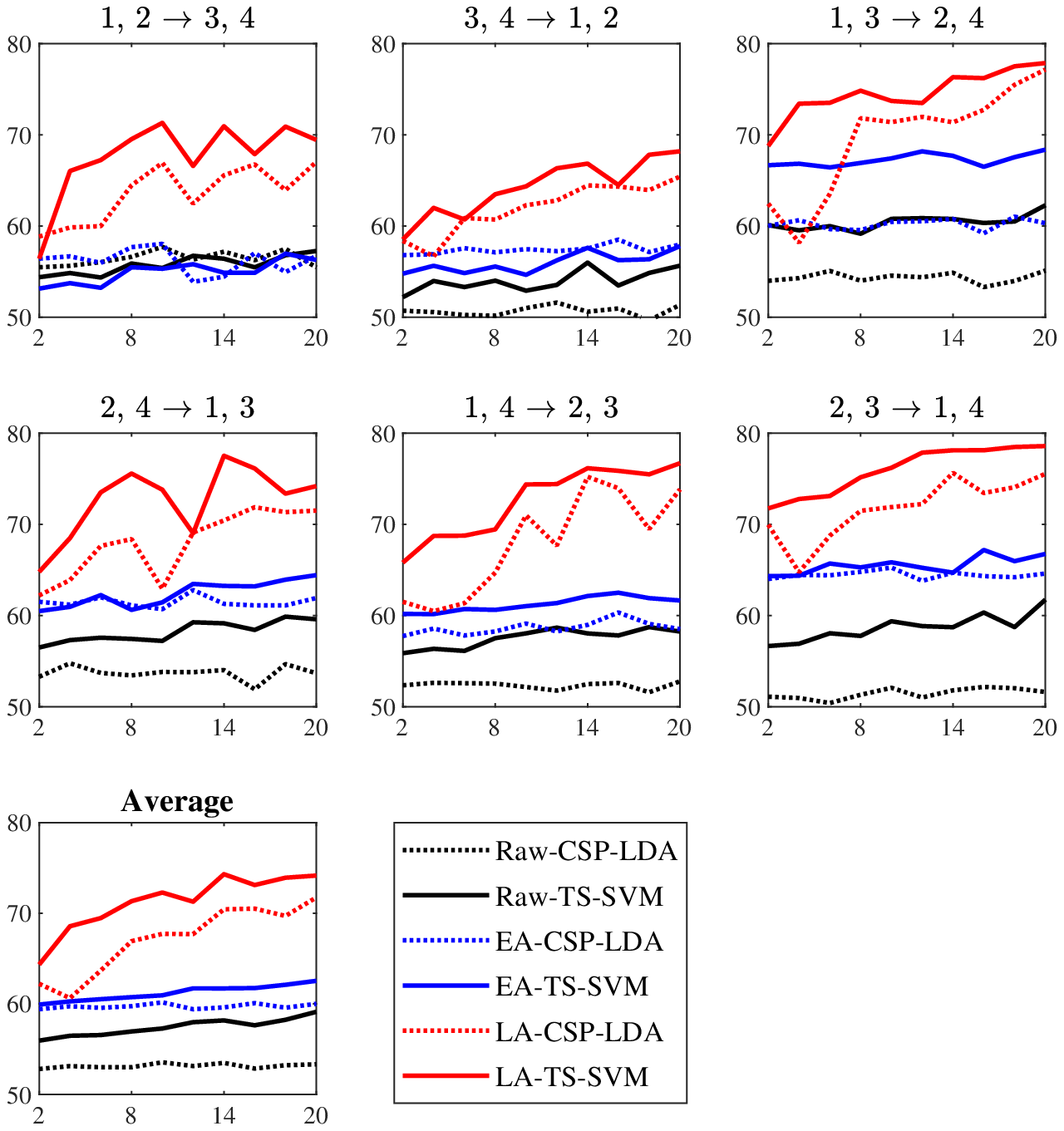}
\caption{Average classification accuracies (\%) in Scenario~II-a. The horizontal axis indicates the number of labeled target subject trials ($k$ in $k$-medoids clustering in Section~\ref{sect:LA}), and the vertical axis the classification accuracies.} \label{fig:2}
\end{figure}

We compared Raw, EA, and LA in the two classification pipelines. Fig.~\ref{fig:2} shows the performances of the six algorithms on the six sub-dataset combinations, and the average. Observe that:
\begin{enumerate}
  \item LA-CSP-LDA always outperformed Raw-CSP-LDA and EA-CSP-LDA, and LA-TS-SVM always outperformed Raw-TS-SVM and EA-TS-SVM. This suggests LA was effective in different feature extraction and classification algorithms.
  \item Comparing the last subfigure in Fig.~\ref{fig:1-a} with the last one in Fig.~\ref{fig:2}, we can observe that the performances of Raw-CSP-LDA and Raw-TS-SVM were lower in Fig.~\ref{fig:2}, which is intuitive, because the label spaces in Scenario~II-a had larger discrepancies. However, the performances of LA-CSP-LDA and LA-TS-SVM did not change much, suggesting that LA can cope well with large label space discrepancies.
\end{enumerate}

For the most extreme case that only one labeled target subject trial from each class is available, the average classification accuracies across the nine subjects in the nine experiments are given in Table~\ref{tab:scenario2}. LA achieved the best performances in all six experiments, regardless of which feature extraction and classification algorithm was used.

\begin{table}[htbp]   \centering \setlength{\tabcolsep}{3mm}
  \caption{Scenario~II-a: average classification accuracies (\%) across the nine subjects when only one labeled target subject trial from each class is available.}
    \begin{tabular}{c|c|ccc} \hline
  Experiment                & Approach&Raw& EA & LA    \\ \hline
   \multirow{2}{*}{1, 2 $\rightarrow$ 3, 4}   &CSP-LDA & 55.48 & 56.42 & \textbf{58.84} \\
     &TS-SVM & 54.38 & 53.13 &\textbf{56.42}   \\   \hline
   \multirow{2}{*}{3, 4 $\rightarrow$ 1, 2}    &CSP-LDA & 50.70 & 56.81 & \textbf{58.37} \\
   &TS-SVM & 52.19 & 54.77 & \textbf{58.53}  \\   \hline
  \multirow{2}{*}{1, 3 $\rightarrow$ 2, 4}     &CSP-LDA & 53.99 & 60.02 & \textbf{62.44}  \\
  &TS-SVM & 60.09 & 66.67 & \textbf{68.78}  \\   \hline
   \multirow{2}{*}{2, 4 $\rightarrow$ 1, 3}    &CSP-LDA & 53.29 & 61.50 & \textbf{62.21} \\
  &TS-SVM & 56.49 & 60.49  & \textbf{64.79}  \\    \hline
   \multirow{2}{*}{1, 4 $\rightarrow$ 2, 3}    &CSP-LDA & 52.35 & 57.75 & \textbf{61.50}  \\
  &TS-SVM & 55.87 & 60.17 & \textbf{65.81}  \\    \hline
  \multirow{2}{*}{2, 3 $\rightarrow$ 1, 4}     &CSP-LDA & 51.10 & 64.01 & \textbf{69.95} \\
  &TS-SVM & 56.65 & 64.32 & \textbf{71.75} \\   \hline
    \end{tabular}   \label{tab:scenario2}
\end{table}

We also performed paired \emph{t}-tests on the AUCs in Fig.~\ref{fig:2}. Each algorithm had $6\times9=54$ AUCs. The $p$-values are shown in Table~\ref{tab:ttest2}, where the statistically significant ones are marked in bold. LA-CSP-LDA significantly outperformed EA-CSP-LDA, and LA-TS-SVM significantly outperformed EA-TS-SVM.

\begin{table}[htpb] \centering \setlength{\tabcolsep}{2.5mm}
\caption{Scenario~II-a: $p$-values of paired $t$-tests on the AUCs of the classification accuracy curves in Fig.~\ref{fig:2}. The null hypothesis was rejected if $p< \alpha$, where $\alpha=0.05$.}   \label{tab:ttest2}
\begin{tabular}{c|cc}   \hline
 & EA-CSP-LDA & EA-TS-SVM    \\  \hline
 LA-CSP-LDA & \textbf{0.0000}& \\ \hline
 LA-TS-SVM & & \textbf{0.0000} \\ \hline
\end{tabular}
\end{table}

\emph{\textbf{Question 2:} Can LA be integrated with other DA approaches to further improve the classification performance?}

\begin{figure}[htpb]\centering
\includegraphics[width=\linewidth,clip]{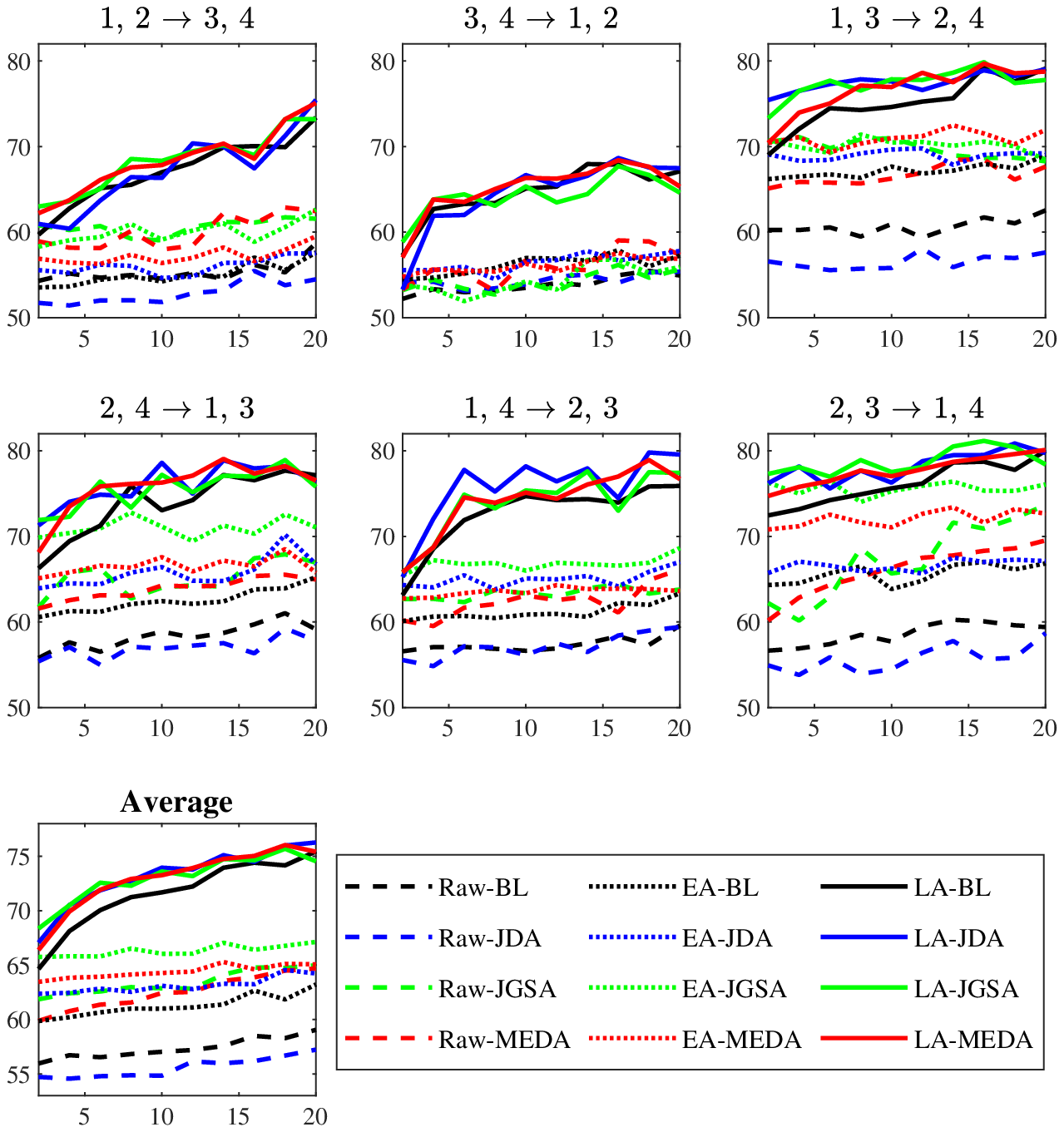}
\caption{Average classification accuracies (\%) in Scenario~II-a, when additional DA approaches were used after LA. The horizontal axis indicates the number of labeled target subject trials ($k$ in $k$-medoids clustering in Section~\ref{sect:LA}), and the vertical axis the classification accuracies.}\label{fig:2_TL}
\end{figure}

Again, we considered the case when there were additional DA approaches after LA. The results are shown in Fig.~\ref{fig:2_TL}. Observe that:
\begin{enumerate}
  \item LA-BL always outperformed Raw-BL and EA-BL, LA-JDA always outperformed Raw-JDA and EA-JDA, LA-JGSA always outperformed Raw-JGSA and EA-JGSA, and LA-MEDA always outperformed Raw-MEDA and EA-MEDA. These suggest that LA was effective regardless of whether an additional DA approach was used or not.
\item LA-BL outperformed Raw-JDA, Raw-JGSA and Raw-MEDA in all six experiments, suggesting that LA can outperform classical DA approaches such as JDA, JGSA and MEDA.
  \item Generally, LA-JDA, LA-JGSA, LA-MEDA outperformed LA-BL, suggesting again that it may be advantageous to integrate an additional DA approach with LA.
\end{enumerate}

The results of paired \emph{t}-tests on the AUCs in Fig.~\ref{fig:2_TL} are shown in Table~\ref{tab:ttest2-TL}, which are consistent with those in the last two subsections: LA significantly outperformed EA and classical DA approaches, and the classification performance can be further significantly improved when LA was integrated with other DA approaches.

\begin{table}[htpb] \centering \setlength{\tabcolsep}{2.5mm}
\caption{Scenario~II-a: $p$-values of paired $t$-tests on the AUCs in Fig.~\ref{fig:2_TL}. The null hypothesis was rejected if $p< \alpha$, where $\alpha=0.05$.}   \label{tab:ttest2-TL}
\begin{tabular}{c|cccc}   \hline
 & LA-BL & EA-JDA &EA-JGSA & EA-MEDA  \\ \hline
 LA-BL & &\textbf{0.0000} &\textbf{0.0000} &\textbf{0.0000}\\ \hline
 LA-JDA &\textbf{0.0109} & \textbf{0.0000} & &\\ \hline
 LA-JGSA &\textbf{0.0116}&&\textbf{0.0000}&\\ \hline
 LA-MEDA &\textbf{0.0009}&&&\textbf{0.0000}\\ \hline
\end{tabular}
\end{table}

\subsection{Scenario~II-b: Same Feature Space and Completely Different Label Spaces in Multi-Class Classification} \label{sect:scenario2-b}

This subsection considers the multi-class classification scenario that the source and the target subjects have the same feature space but completely different label spaces.

Ideally, if Dataset~2a has six or more different classes, we can perform studies like ``$1, 2, 3\rightarrow4,5,6$" in multi-class classification. Unfortunately, Dataset~2a only has four classes. So, we mismatched the labels between the target and the source subjects, to simulate completely different label spaces in multi-class classification.

Assume the labels of the source subjects are `1', `2' and `3', and the labels of the target subject are `1', `2' and `4'. Then, we match `1' of the source subjects with `2' of the target subject, `2' of the source subjects with `1' of the target subject, and `3' of the source subjects with `4' of the target subject, i.e., `$1,2,3\rightarrow2,1,4$. A potential application scenario of this setting is that for the source subjects we know which trials belong to the same class, but do not know the specific class labels. So, we randomly match them to the labels of the target subject.

\emph{\textbf{Question 1:} Can LA be used as an effective preprocessing step before different feature extraction and classification algorithms?}

\begin{figure*}[htpb]\centering
\includegraphics[width=\linewidth,clip]{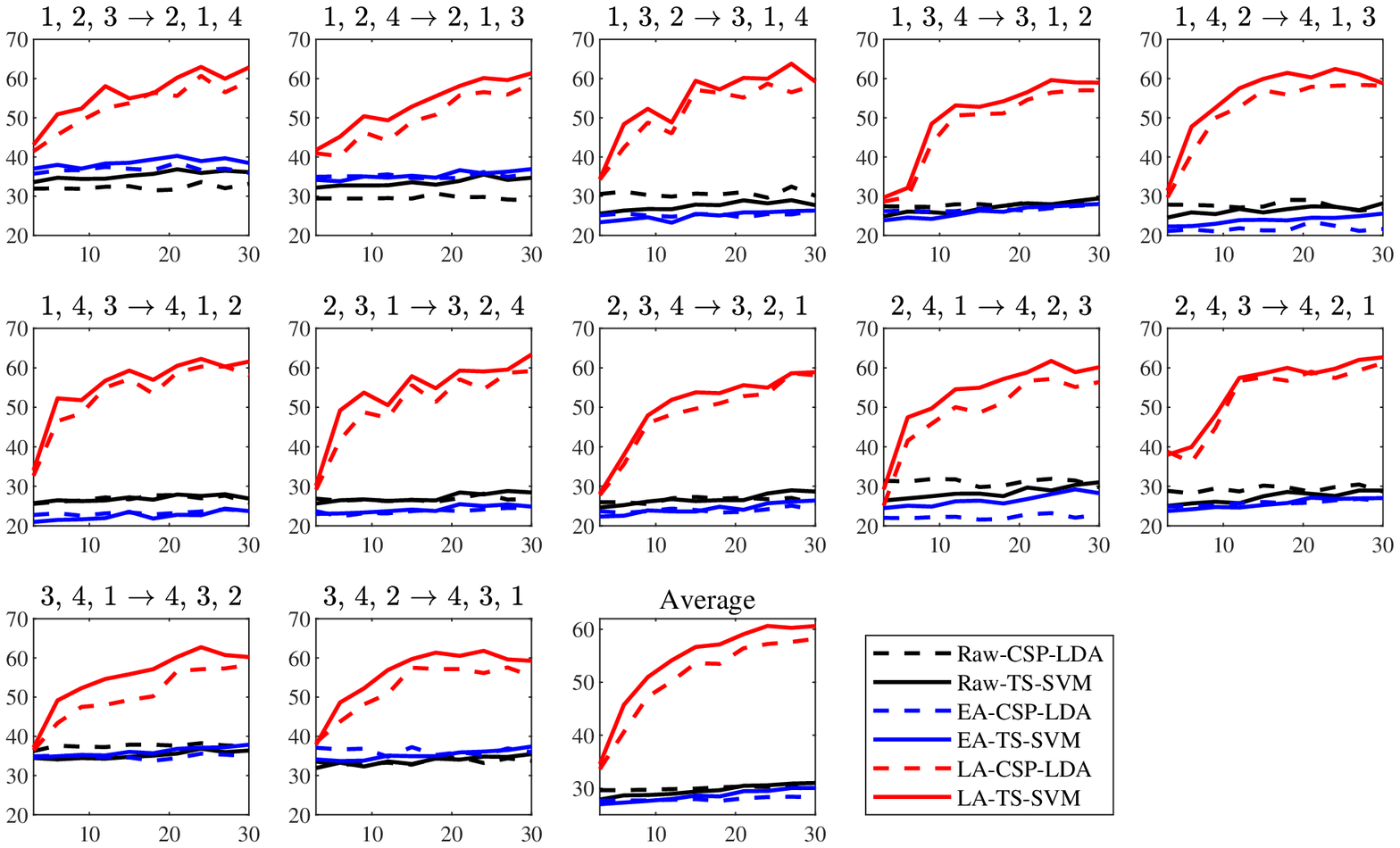}
\caption{Average classification accuracies (\%) in Scenario~II-b. The horizontal axis indicates the number of labeled target subject trials ($k$ in $k$-medoids clustering in Section~\ref{sect:LA}), and the vertical axis the classification accuracies.} \label{fig:2-b}
\end{figure*}

Again, we compared Raw, EA, and LA in the two classification pipelines to answer this question. Fig.~\ref{fig:2-b} shows the performances of the six algorithms on 12 different sub-dataset combinations, where each subfigure shows the average classification accuracies across the nine subjects (each as the target subject once). The last subfigure shows the average performances across the 12 experiments. The title of each subfigure shows the sub-datasets used, and also how we matched the labels between the two sub-datasets. Observe that:
\begin{enumerate}
  \item LA-CSP-LDA always outperformed Raw-CSP-LDA and EA-CSP-LDA, and LA-TS-SVM always outperformed Raw-TS-SVM and EA-TS-SVM. This suggests that  was effective in different feature extraction and classification algorithms.
  \item LA in Fig.~\ref{fig:2-b} achieved much larger performance improvements over Raw and EA than those in Fig.~\ref{fig:1-b}. When the labels mismatched, the algorithms without LA (i.e., Raw-CSP-LDA, Raw-TS-SVM, EA-CSP-LDA and EA-TS-SVM) performed very poorly. However, the performances of LA-CSP-LDA and LA-TS-SVM were very consistent, suggesting that LA can cope well with large label space discrepancies.
\end{enumerate}

Paired \emph{t}-tests on the AUCs in Fig.~\ref{fig:2-b} were also performed to check if the differences between different algorithms were statistically significant. The results are shown in Table~\ref{tab:ttest2-b}, which indicate that LA-CSP-LDA significantly outperformed EA-CSP-LDA, and LA-TS-SVM significantly outperformed EA-TS-SVM.

\begin{table}[htpb] \centering \setlength{\tabcolsep}{2.5mm}
\caption{Scenario~II-b: $p$-values of paired $t$-tests on the AUCs of the classification accuracy curves in Fig.~\ref{fig:2-b}. The null hypothesis was rejected if $p< \alpha$, where $\alpha=0.05$.}   \label{tab:ttest2-b}
\begin{tabular}{c|cc}   \hline
 & EA-CSP-LDA & EA-TS-SVM  \\ \hline
 LA-CSP-LDA & \textbf{0.0000}&\\ \hline
 LA-TS-SVM & & \textbf{0.0000} \\ \hline
\end{tabular}
\end{table}

\emph{\textbf{Question 2:} Can LA be integrated with other DA approaches to further improve the classification performance?}

\begin{figure*}[htpb]\centering
\includegraphics[width=\linewidth,clip]{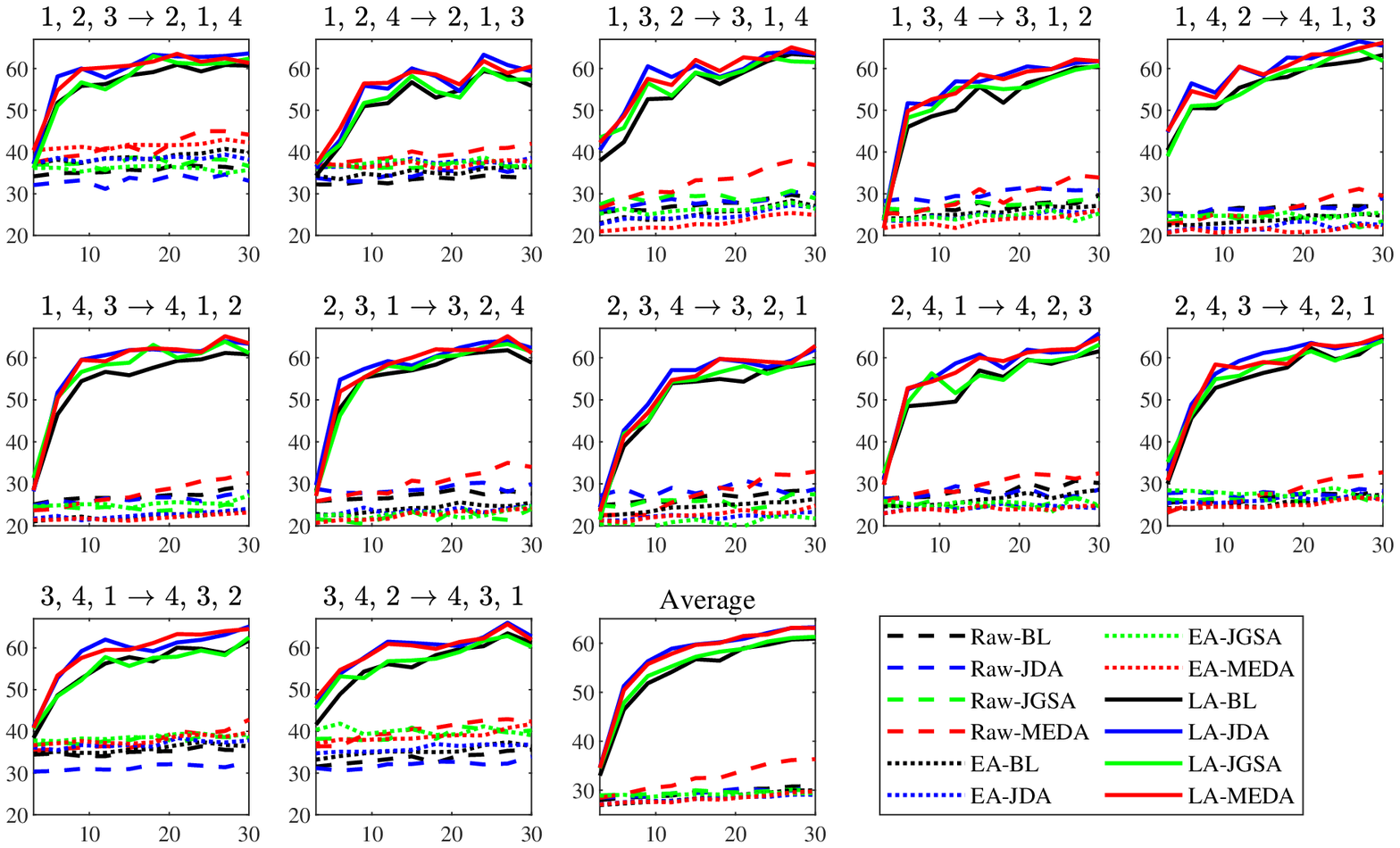}
\caption{Average classification accuracies (\%) in Scenario~II-b, when other DA approaches were used after LA. The horizontal axis indicates the number of labeled target subject trials ($k$ in $k$-medoids clustering in Section~\ref{sect:LA}), and the vertical axis the classification accuracies.} \label{fig:2TL-b}
\end{figure*}

Again, we combined Raw, EA, LA with different DA approaches and obtained 12 algorithms to be compared. Fig.~\ref{fig:2TL-b} shows their performances on the 12 sub-dataset combinations, and the average across the 12 experiments. Observe that:
\begin{enumerate}
  \item LA-BL always outperformed Raw-BL and EA-BL, LA-JDA always outperformed Raw-JDA and EA-JDA, LA-JGSA always outperformed Raw-JGSA and EA-JGSA, and LA-MEDA always outperformed Raw-MEDA and EA-MEDA. These suggest that LA was effective regardless of whether additional DA approaches were used or not.
  \item LA-BL always outperformed Raw-JDA, Raw-JGSA and Raw-MEDA, suggesting that LA can outperform classical DA approaches such as JDA, JGSA and MEDA.
  \item Generally, LA-JDA, LA-JGSA and LA-MEDA outperformed LA-BL, suggesting that it may be advantageous to integrate additional DA approaches with LA.
  \item When the labels were mismatched, the algorithms without LA performed very poorly. However, the algorithms with LA performed consistently good, suggesting that LA can cope well with large label space discrepancies.
\end{enumerate}

Table~\ref{tab:ttest2-b-TL} shows the results of paired \emph{t}-tests on the AUCs in Fig. \ref{fig:2TL-b}. The conclusions in binary classification still hold in multi-class classification: LA significantly outperformed EA and classical DA approaches, and its performance can be further significantly improved when integrated with other DA approaches.

\begin{table}[htpb] \centering \setlength{\tabcolsep}{2.5mm}
\caption{Scenario~II-b: $p$-values of paired $t$-tests on the AUCs in Fig.~\ref{fig:2TL-b}. The null hypothesis was rejected if $p< \alpha$, where $\alpha=0.05$.}   \label{tab:ttest2-b-TL}
\begin{tabular}{c|cccc}   \hline
 & LA-BL & EA-JDA &EA-JGSA & EA-MEDA  \\  \hline
 LA-BL & &\textbf{0.0000} &\textbf{0.0000} &\textbf{0.0000}\\ \hline
 LA-JDA &\textbf{0.0000} & \textbf{0.0000} & &\\ \hline
 LA-JGSA &\textbf{0.0009}&&\textbf{0.0000}&\\ \hline
 LA-MEDA &\textbf{0.0000}&&&\textbf{0.0000}\\ \hline
\end{tabular}
\end{table}

\subsection{Scenario III: Different Feature Spaces and Different Label Spaces}\label{sect:scenario3}

This subsection considers the most challenging scenario: the source and target subjects have different feature spaces and also completely different label spaces. We chose ``Classes 3, 4" (``feet" and ``tongue") from Dataset~2a as the target dataset, and Dataset~1 as the source dataset. Each time we picked one subject from ``Classes 3, 4" as the target subject, and all seven subjects from Dataset~1 as the source subjects. In this scenario, the source dataset and target dataset were collected from different EEG headsets with different numbers of channels at different locations, so they had different feature spaces. In addition, for Dataset~1, Subjects~1 and 6 performed ``left hand" and ``feet" tasks, whereas other subjects performed ``left hand" and ``right hand" tasks. So, the source and target subjects also had partially or completely different label spaces.

\emph{\textbf{Question 1:} Can LA be used as an effective preprocessing step before different feature extraction and classification algorithms?}

\begin{figure}[htpb]\centering
\includegraphics[width=\linewidth,clip]{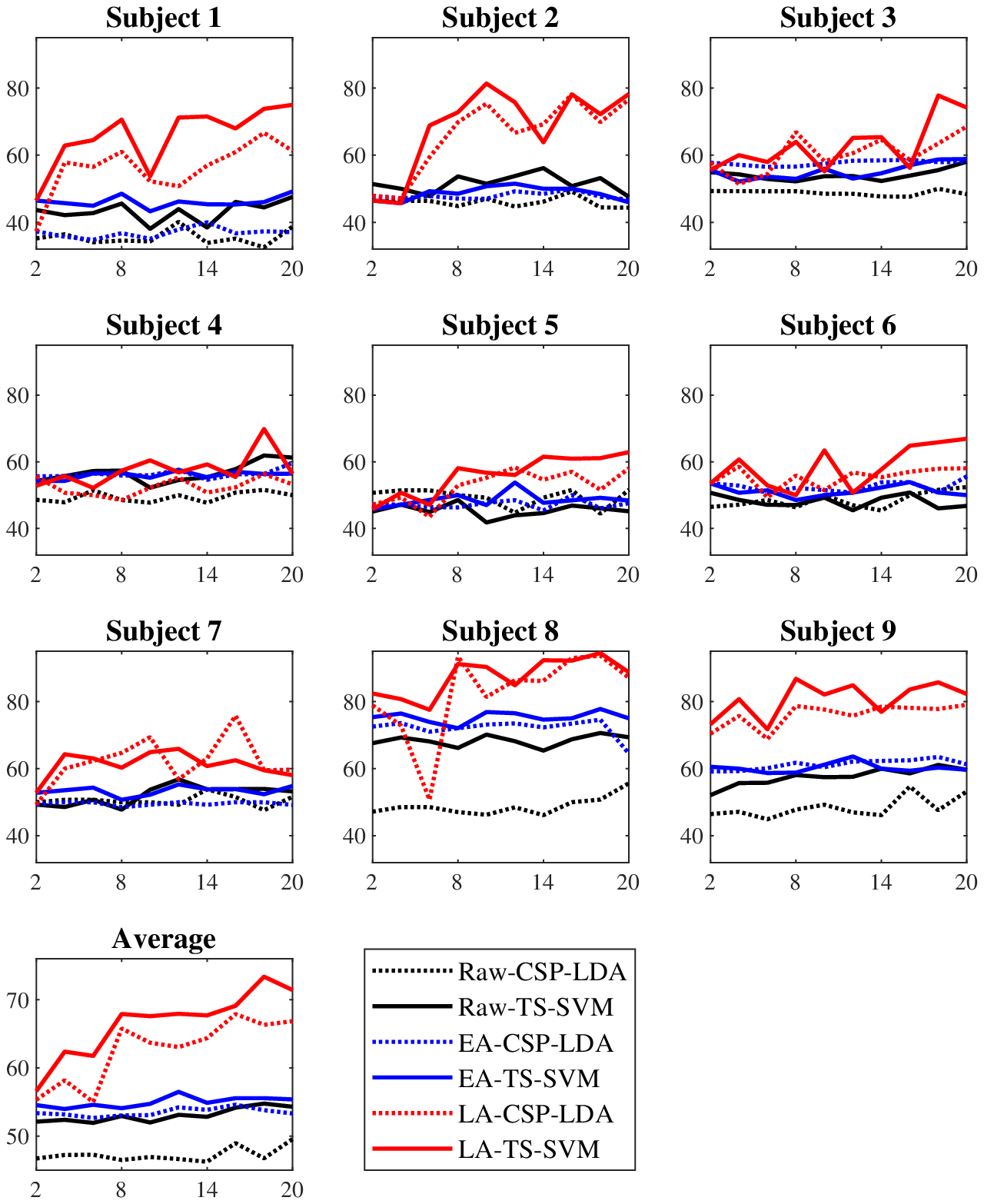}
\caption{Average classification accuracies (\%) in Scenario~III. The horizontal axis indicates the number of labeled target subject trials ($k$ in $k$-medoids clustering in Section~\ref{sect:LA}), and the vertical axis the classification accuracies.}\label{fig:3}
\end{figure}

We selected the source EEG channels as those closest to the target EEG channels \cite{drwuTNSRE2016}, and compared different algorithms. Fig.~\ref{fig:3} shows the experimental results when LA was used before different feature extraction and classification algorithms, and Table~\ref{tab:ttest3} shows the $p$-values of paired \emph{t}-tests on the AUCs. LA-CSP-LDA significantly outperformed EA-CSP-LDA, and LA-TS-SVM significantly outperformed EA-TS-SVM. These suggest that LA was effective in different feature extraction and classification algorithms.

\begin{table}[htpb] \centering \setlength{\tabcolsep}{2mm}
\caption{Scenario~III: $p$-values of the paired $t$-tests on the AUCs of the accuracy curves in Fig.~\ref{fig:3}. The null hypothesis was rejected if $p< \alpha$, where $\alpha=0.05$.}   \label{tab:ttest3}
\begin{tabular}{c|cc}   \hline
 & EA-CSP-LDA & EA-TS-SVM   \\  \hline
 LA-CSP-LDA & \textbf{0.0082}&\\ \hline
 LA-TS-SVM & & \textbf{0.0006} \\ \hline
\end{tabular}
\end{table}

\emph{\textbf{Question 2:} Can LA be integrated with other DA approaches to further improve the classification performance?}

\begin{figure}[htpb]\centering
\includegraphics[width=\linewidth,clip]{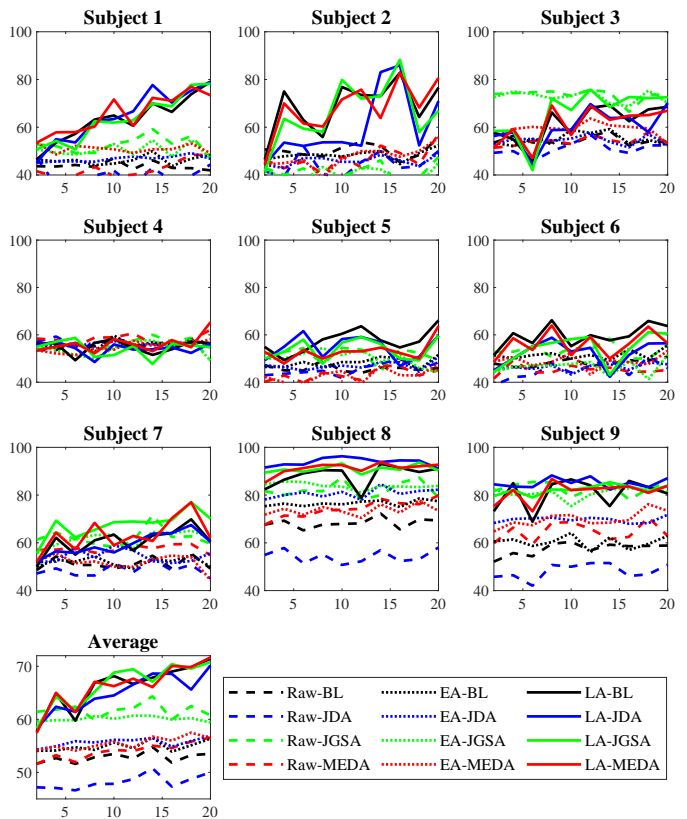}
\caption{Average classification accuracies (\%) in Scenario~III, when additional DA approaches were used after LA. The horizontal axis indicates the number of labeled target subject trials ($k$ in $k$-medoids clustering in Section~\ref{sect:LA}), and the vertical axis the classification accuracies.}\label{fig:3TL}
\end{figure}

Fig.~\ref{fig:3TL} shows the experimental results with and without additional DA approaches. Generally, LA was effective regardless of whether additional DA approaches were used or not. Table~\ref{tab:ttest3-TL} shows the $p$-values of paired \emph{t}-tests on the AUCs in Fig.~\ref{fig:3TL}. LA-BL significantly outperformed EA-JDA and EA-MEDA, LA-JDA significantly outperformed EA-JDA, and LA-MEDA significantly outperformed EA-MEDA. However, unlike before, the integration of LA with other DA approaches did not significantly outperform LA-BL. Nevertheless, LA did not degrade the performance of these DA approaches, either.

\begin{table}[htpb] \centering \setlength{\tabcolsep}{2.5mm}
\caption{Scenario~III: $p$-values of the paired $t$-tests on the AUCs of the accuracy curves in Fig.~\ref{fig:3TL}. The null hypothesis was rejected if $p< \alpha$, where $\alpha=0.05$.}   \label{tab:ttest3-TL}
\begin{tabular}{c|cccc}   \hline
 & LA-BL & EA-JDA &EA-JGSA & EA-MEDA  \\  \hline
 LA-BL & &\textbf{0.0017} &0.1335 &\textbf{0.0011}\\ \hline
 LA-JDA &0.3733&\textbf{0.0026}  & &\\ \hline
 LA-JGSA &0.8449&&0.0691&\\ \hline
 LA-MEDA &0.9777&&&\textbf{0.0012}\\ \hline
\end{tabular}
\end{table}

\subsection{Computational Complexity of LA}

The time complexity of LA is $O({N_T}^2)$, where $N_T$ is the number of target domain trials. The most time-consuming operation in LA is $k$-medoids clustering.

We also empirically evaluated the computational cost of LA in Scenario~III, by comparing Raw-TS-SVM and LA-TS-SVM. The platform was a Lenovo ThinkPad laptop with Intel Core i5-6200U CPU@2.30GHz, 4GB memory, and 190 GB SSD, running 64-bit Windows 10 and Matlab 2018b. The results are shown in Table~\ref{tab:time}, which were averaged across different numbers of labeled target trials (from 2 to 20) and nine target subjects. LA only increased the computing time very slightly.

\begin{table}[htpb]\centering  \setlength{\tabcolsep}{6mm}
\caption{The computing time (seconds) of Raw-TS-SVM and LA-TS-SVM in Scenario~III.}\label{tab:time}
\begin{tabular}{c|cc}   \hline
 & mean & std   \\  \hline
 Raw-TS-SVM & 2.1963&0.1492\\ \hline
 LA-TS-SVM & 2.3669& 0.3469 \\ \hline
\end{tabular}
\end{table}

\section{Conclusions and Future Research} \label{sect:conclusions}

Transfer learning, or domain adaptation, has been successfully used to reduce the subject-specific calibration effort in BCIs. However, most existing DA approaches require the source subjects share the same feature space and also the same label space as the target subject, which may not always hold in real-world applications. This paper has proposed a simple yet effective LA approach to cope with different label spaces. Our experiments demonstrated that: 1) LA only needs as few as one labeled sample from each class of the target subject; 2) LA can be used as a preprocessing step before different feature extraction and classification algorithms; and, 3) LA can be integrated with other DA approaches to achieve even better classification performance.

The current LA may still have some limitations, which will be addressed in our future research:
\begin{enumerate}
\item The estimation of each class mean in the target domain is very important to the performance of LA. Currently LA uses $k$-medoids clustering to select a few trials to label, which could be improved.
\item LA copes well with different labels spaces, but does not pay special attention to different feature spaces (although it can also be used in this case). This may explain why there were relatively less performance improvements when integrated with other DA approaches in Scenario~III. We will specifically consider different feature spaces in the future.
\item The current LA approach was specifically designed for EEG trials, and uses 2D covariance matrices as the input features. We will extend it to 1D features so that it can have broader applications in other domains.
\end{enumerate}

\section*{Acknowledgment}

This research was supported by the National Natural Science Foundation of China under Grant 61873321 and Hubei Technology Innovation Platform under Grant 2019AEA171.

% Generated by IEEEtran.bst, version: 1.14 (2015/08/26)


\begin{thebibliography}{10}
\providecommand{\url}[1]{#1}
\csname url@samestyle\endcsname
\providecommand{\newblock}{\relax}
\providecommand{\bibinfo}[2]{#2}
\providecommand{\BIBentrySTDinterwordspacing}{\spaceskip=0pt\relax}
\providecommand{\BIBentryALTinterwordstretchfactor}{4}
\providecommand{\BIBentryALTinterwordspacing}{\spaceskip=\fontdimen2\font plus
\BIBentryALTinterwordstretchfactor\fontdimen3\font minus
  \fontdimen4\font\relax}
\providecommand{\BIBforeignlanguage}[2]{{%
\expandafter\ifx\csname l@#1\endcsname\relax
\typeout{** WARNING: IEEEtran.bst: No hyphenation pattern has been}%
\typeout{** loaded for the language `#1'. Using the pattern for}%
\typeout{** the default language instead.}%
\else
\language=\csname l@#1\endcsname
\fi
#2}}
\providecommand{\BIBdecl}{\relax}
\BIBdecl

\bibitem{Wolpaw2002}
J.~R. Wolpaw, N.~Birbaumer, D.~J. McFarland, G.~Pfurtscheller, and T.~M.
  Vaughan, ``Brain-computer interfaces for communication and control,''
  \emph{Clinical Neurophysiology}, vol. 113, no.~6, pp. 767--791, 2002.

\bibitem{Lance2012}
B.~J. Lance, S.~E. Kerick, A.~J. Ries, K.~S. Oie, and K.~McDowell,
  ``Brain-computer interface technologies in the coming decades,'' \emph{Proc.
  of the {IEEE}}, vol. 100, no.~3, pp. 1585--1599, 2012.

\bibitem{Koles1990}
Z.~J. Koles, M.~S. Lazar, and S.~Z. Zhou, ``Spatial patterns underlying
  population differences in the background {EEG},'' \emph{Brain Topography},
  vol.~2, no.~4, pp. 275--284, 1990.

\bibitem{Muller1999}
J.~M{\"u}ller-Gerking, G.~Pfurtscheller, and H.~Flyvbjerg, ``Designing optimal
  spatial filters for single-trial {EEG} classification in a movement task,''
  \emph{Clinical Neurophysiology}, vol. 110, no.~5, pp. 787--798, 1999.

\bibitem{Ramoser2000}
H.~Ramoser, J.~Muller-Gerking, and G.~Pfurtscheller, ``Optimal spatial
  filtering of single trial {EEG} during imagined hand movement,'' \emph{{IEEE}
  Trans. on Rehabilitation Engineering}, vol.~8, no.~4, pp. 441--446, 2000.

\bibitem{he2018spatial}
H.~He and D.~Wu, ``Spatial filtering for brain computer interfaces: A
  comparison between the common spatial pattern and its variant,'' in
  \emph{Proc. {IEEE} Int'l. Conf. on Signal Processing, Communications and
  Computing}, Qingdao, China, Sep. 2018, pp. 1--6.

\bibitem{Barachant2012}
A.~Barachant, S.~Bonnet, M.~Congedo, and C.~Jutten, ``Multiclass brain-computer
  interface classification by {R}iemannian geometry,'' \emph{{IEEE} Trans. on
  Biomedical Engineering}, vol.~59, no.~4, pp. 920--928, 2012.

\bibitem{Yger2017}
F.~Yger, M.~Berar, and F.~Lotte, ``Riemannian approaches in brain-computer
  interfaces: a review,'' \emph{{IEEE} Trans. on Neural Systems and
  Rehabilitation Engineering}, vol.~25, no.~10, pp. 1753--1762, 2017.

\bibitem{He2015}
B.~He, B.~Baxter, B.~J. Edelman, C.~C. Cline, and W.~W. Ye, ``Noninvasive
  brain-computer interfaces based on sensorimotor rhythms,'' \emph{Proc. of the
  {IEEE}}, vol. 103, no.~6, pp. 907--925, 2015.

\bibitem{Pfurtscheller2008}
G.~Pfurtscheller, G.~R. M{\"u}ller-Putz, R.~Scherer, and C.~Neuper,
  ``Rehabilitation with brain-computer interface systems,'' \emph{Computer},
  vol.~41, no.~10, pp. 58--65, 2008.

\bibitem{Nicolas-Alonso2012}
L.~F. Nicolas-Alonso and J.~Gomez-Gil, ``Brain computer interfaces, a review,''
  \emph{Sensors}, vol.~12, no.~2, pp. 1211--1279, 2012.

\bibitem{Erp2012}
J.~van Erp, F.~Lotte, and M.~Tangermann, ``Brain-computer interfaces: Beyond
  medical applications,'' \emph{Computer}, vol.~45, no.~4, pp. 26--34, 2012.

\bibitem{Jayaram2016}
V.~Jayaram, M.~Alamgir, Y.~Altun, B.~Scholkopf, and M.~Grosse-Wentrup,
  ``Transfer learning in brain-computer interfaces,'' \emph{{IEEE}
  Computational Intelligence Magazine}, vol.~11, no.~1, pp. 20--31, 2016.

\bibitem{drwuTNSRE2016}
D.~Wu, V.~J. Lawhern, W.~D. Hairston, and B.~J. Lance, ``Switching {EEG}
  headsets made easy: {Reducing} offline calibration effort using active
  weighted adaptation regularization,'' \emph{{IEEE} Trans. on Neural Systems
  and Rehabilitation Engineering}, vol.~24, no.~11, pp. 1125--1137, 2016.

\bibitem{drwuTFS2017}
D.~Wu, V.~J. Lawhern, S.~Gordon, B.~J. Lance, and C.-T. Lin, ``Driver
  drowsiness estimation from {EEG} signals using online weighted adaptation
  regularization for regression ({OwARR}),'' \emph{{IEEE} Trans. on Fuzzy
  Systems}, vol.~25, no.~6, pp. 1522--1535, 2017.

\bibitem{drwuTHMS2017}
D.~Wu, ``Online and offline domain adaptation for reducing {BCI} calibration
  effort,'' \emph{{IEEE} Trans. on Human-Machine Systems}, vol.~47, no.~4, pp.
  550--563, 2017.

\bibitem{drwuSMC2017}
D.~Wu, ``Active semi-supervised transfer learning ({ASTL}) for offline {BCI}
  calibration,'' in \emph{Proc. {IEEE} Int'l. Conf. on Systems, Man and
  Cybernetics}, Banff, Canada, October 2017.

\bibitem{drwuTLCSP2017}
H.~He and D.~Wu, ``Transfer learning enhanced common spatial pattern filtering
  for brain computer interfaces ({BCIs}): Overview and a new approach,'' in
  \emph{Proc. 24th Int'l. Conf. on Neural Information Processing}, Guangzhou,
  China, November 2017.

\bibitem{Kang2009}
H.~Kang, Y.~Nam, and S.~Choi, ``Composite common spatial pattern for
  subject-to-subject transfer,'' \emph{Signal Processing Letters}, vol.~16,
  no.~8, pp. 683--686, 2009.

\bibitem{Lotte2010}
F.~Lotte and C.~Guan, ``Learning from other subjects helps reducing
  brain-computer interface calibration time,'' in \emph{Proc. {IEEE} Int'l.
  Conf. on Acoustics Speech and Signal Processing}, Dallas, TX, March 2010.

\bibitem{Pan2010}
S.~J. Pan and Q.~Yang, ``A survey on transfer learning,'' \emph{{IEEE} Trans.
  on Knowledge and Data Engineering}, vol.~22, no.~10, pp. 1345--1359, 2010.

\bibitem{long2013transfer}
M.~Long, J.~Wang, G.~Ding, J.~Sun, and P.~S. Yu, ``Transfer feature learning
  with joint distribution adaptation,'' in \emph{Proc. {IEEE} Int'l. Conf. on
  Computer Vision}, Sydney, Australia, Dec. 2013, pp. 2200--2207.

\bibitem{zhang2017joint}
J.~Zhang, W.~Li, and P.~Ogunbona, ``Joint geometrical and statistical alignment
  for visual domain adaptation,'' in \emph{Proc. {IEEE} Conf. on Computer
  Vision and Pattern Recognition}, Honolulu, HI, Jul. 2017, pp. 1859--1867.

\bibitem{wang2018visual}
J.~Wang, W.~Feng, Y.~Chen, H.~Yu, M.~Huang, and P.~S. Yu, ``Visual domain
  adaptation with manifold embedded distribution alignment,'' in \emph{Proc.
  26th ACM Int'l Conf. on Multimedia}, Seoul, South Korea, Oct. 2018, pp.
  402--410.

\bibitem{Lu2015Transfer}
J.~Lu, V.~Behbood, P.~Hao, H.~Zuo, S.~Xue, and G.~Zhang, ``Transfer learning
  using computational intelligence: a survey,'' \emph{Knowledge-Based Systems},
  vol.~80, pp. 14--23, 2015.

\bibitem{Zanini2018}
P.~Zanini, M.~Congedo, C.~Jutten, S.~Said, and Y.~Berthoumieu, ``Transfer
  learning: a {R}iemannian geometry framework with applications to
  brain-computer interfaces,'' \emph{{IEEE} Trans. on Biomedical Engineering},
  vol.~65, no.~5, pp. 1107--1116, 2018.

\bibitem{he2019transfer}
H.~He and D.~Wu, ``Transfer learning for brain-computer interfaces: A
  {Euclidean} space data alignment approach,'' \emph{{IEEE} Trans. on
  Biomedical Engineering}, vol.~67, no.~2, pp. 399--410, 2020.

\bibitem{day2017survey}
O.~Day and T.~M. Khoshgoftaar, ``A survey on heterogeneous transfer learning,''
  \emph{Journal of Big Data}, vol.~4, no.~1, p.~29, 2017.

\bibitem{liu2018unsupervised}
F.~Liu, J.~Lu, and G.~Zhang, ``Unsupervised heterogeneous domain adaptation via
  shared fuzzy equivalence relations,'' \emph{{IEEE} Trans. on Fuzzy Systems},
  vol.~26, no.~6, pp. 3555--3568, 2018.

\bibitem{panareda2017open}
P.~P. Busto and J.~Gall, ``Open set domain adaptation,'' in \emph{Proc. {IEEE}
  Int'l Conf. on Computer Vision}, Venice, Italy, Oct. 2017, pp. 754--763.

\bibitem{saito2018open}
K.~Saito, S.~Yamamoto, Y.~Ushiku, and T.~Harada, ``Open set domain adaptation
  by backpropagation,'' in \emph{Proc. European Conf. on Computer Vision},
  Munich, Germany, Sep. 2018, pp. 153--168.

\bibitem{fang2019open}
Z.~Fang, J.~Lu, F.~Liu, J.~Xuan, and G.~Zhang, ``Open set domain adaptation:
  Theoretical bound and algorithm,'' \emph{arXiv preprint arXiv:1907.08375},
  2019.

\bibitem{you2019universal}
K.~You, M.~Long, Z.~Cao, J.~Wang, and M.~I. Jordan, ``Universal domain
  adaptation,'' in \emph{Proc. {IEEE} Conf. on Computer Vision and Pattern
  Recognition}, Long Beach, CA, Jun. 2019, pp. 2720--2729.

\bibitem{he2019channel}
H.~He and D.~Wu, ``Channel and trials selection for reducing covariate shift in
  {EEG}-based brain-computer interfaces,'' in \emph{2019 {IEEE} Int'l Conf. on
  Systems, Man and Cybernetics}, Bari, Italy, Oct. 2019, pp. 3635--3640.

\bibitem{gretton2007kernel}
A.~Gretton, K.~Borgwardt, M.~Rasch, B.~Sch{\"o}lkopf, and A.~J. Smola, ``A
  kernel method for the two-sample-problem,'' in \emph{Proc. Advances in Neural
  Information Processing Systems}, Vancouver, Canada, Dec. 2007, pp. 513--520.

\bibitem{sun2016return}
B.~Sun, J.~Feng, and K.~Saenko, ``Return of frustratingly easy domain
  adaptation,'' in \emph{Proc. 30th {AAAI} Conf. on Artificial Intelligence},
  vol.~6, no.~7, Phoenix, AZ, Feb. 2016, pp. 2058--2065.

\bibitem{shimodaira2000improving}
H.~Shimodaira, ``Improving predictive inference under covariate shift by
  weighting the log-likelihood function,'' \emph{Journal of Statistical
  Planning and Inference}, vol.~90, no.~2, pp. 227--244, 2000.

\bibitem{sugiyama2007direct}
M.~Sugiyama, S.~Nakajima, H.~Kashima, P.~von B{\"{u}}nau, and M.~Kawanabe,
  ``Direct importance estimation with model selection and its application to
  covariate shift adaptation,'' in \emph{Proc. 21th Annual Conf. on Neural
  Information Processing Systems}, Vancouver, Canada, Dec. 2007, pp.
  1433--1440.

\bibitem{utgoff1986shift}
P.~E. Utgoff, \emph{Machine learning: An artificial intelligence
  approach}.\hskip 1em plus 0.5em minus 0.4em\relax CA: Morgan Kaufmann, 1986,
  vol.~2, ch. Shift of bias for inductive concept learning, pp. 107--148.

\bibitem{arsigny2007geometric}
V.~Arsigny, P.~Fillard, X.~Pennec, and N.~Ayache, ``Geometric means in a novel
  vector space structure on symmetric positive-definite matrices,''
  \emph{{SIAM} Journal on Matrix Analysis and Applications}, vol.~29, no.~1,
  pp. 328--347, 2007.

\bibitem{Blankertz2007}
B.~Blankertz, G.~Dornhege, M.~Krauledat, K.~R. Muller, and G.~Curio, ``The
  non-invasive {B}erlin brain-computer interface: Fast acquisition of effective
  performance in untrained subjects,'' \emph{NeuroImage}, vol.~37, no.~2, pp.
  539--550, 2007.

\bibitem{Delorme2004}
A.~Delorme and S.~Makeig, ``{EEGLAB}: an open source toolbox for analysis of
  single-trial {EEG} dynamics including independent component analysis,''
  \emph{Journal of Neuroscience Methods}, vol. 134, pp. 9--21, 2004.

\bibitem{Blankertz2008}
B.~Blankertz, R.~Tomioka, S.~Lemm, M.~Kawanabe, and K.~R. Muller, ``Optimizing
  spatial filters for robust {EEG} single-trial analysis,'' \emph{{IEEE} Signal
  Processing Magazine}, vol.~25, no.~1, pp. 41--56, 2008.

\bibitem{Maaten2008}
L.~van~der Maaten and G.~Hinton, ``Visualizing data using t-{SNE},''
  \emph{Journal of Machine Learning Research}, vol.~9, pp. 2579--2605, 2008.

\bibitem{Dornhege2004}
G.~Dornhege, G.~C. B.~Blankertz, and K.-R. Muller, ``Boosting bit rates in
  non-invasive {EEG} single-trial classifications by feature combination and
  multi-class paradigms,'' \emph{{IEEE} Trans. on Biomedical Engineering},
  vol.~51, no.~6, pp. 993--1002, 2004.

\end{thebibliography}
\end{document}